\documentclass[final,1p,times]{elsarticle}

\usepackage{graphicx}

\usepackage{amssymb}
\usepackage{amsthm}

\usepackage{amsmath,url}

\usepackage{algorithm}
\usepackage{algorithmic}

  \newcommand{\abs}[2][]{\ensuremath{\left|#2\right|}}

  \newcommand{\pad}[2][]{\ensuremath{\frac{\partial #1}{\partial #2}}}

  \newcommand{\R}{\mathbb{R}}

\newtheorem{definition}{Definition}
\newtheorem{problem}{Problem}

\journal{Computer Physics Communications}

\begin{document}

\begin{frontmatter}


\title{Curvilinear Grids for WENO Methods in Astrophysical Simulations}

\author[fmv,mpa]{H.~Grimm-Strele\corref{cor1}}
\ead{hannes.grimm-strele@univie.ac.at} 
\author[fmv]{F.~Kupka} 
\author[fmv]{H.~J.~Muthsam}

\cortext[cor1]{corresponding author}
\address[fmv]{Institute of Mathematics, University of Vienna, Nordbergstra{\ss}e~15,
                       A-1090~Vienna, Austria}
\address[mpa]{Max--Planck Institute for Astrophysics, Karl-Schwarzschild-Strasse~1, 
                       85741~Garching, Germany}

\begin{abstract}
We investigate the applicability of curvilinear grids in the context of astrophysical 
simulations and WENO schemes. With the non-smooth mapping functions from 
\citet{Calhounetal2008}, we can tackle many astrophysical problems which were out of 
scope with the standard grids in numerical astrophysics.
We describe the difficulties occurring when implementing curvilinear coordinates
into our WENO code, and how we overcome them. We illustrate the theoretical results 
with numerical data. 
The WENO finite difference scheme works only for high Mach number flows and 
smooth mapping functions whereas the finite volume scheme gives accurate results 
even for low Mach number flows and on non-smooth grids.
\end{abstract}


\end{frontmatter}

\section{Introduction}
\label{sec-intro}

There are many astrophysical applications where the physical domain of 
interest is a sphere or a circle, e.g.\ the numerical simulation of core
convection \citep{BrowningBrunToomre2004,CaiChanDeng2011} or of convection 
in giant planets \citep{EvonukGlatzmaier2006,EvonukGlatzmaier2007}. The 
usability of spherical coordinate systems is restricted due to the grid 
singularity in the centre of the sphere as discussed, for instance, 
by \citet{EvonukGlatzmaier2007}. With a Cartesian grid, a huge part of 
the computational resources are wasted \citep{FreytagSteffenDorch2002} and 
to improve resolution along spheres, complex adaptive mesh refinements have 
to be used \citep{ZingaleNonakaAlmgrenetal2013} to keep the computational 
requirements manageable. Therefore, all default grids used in numerical 
astrophysics have some fatal deficiencies.

For the specific case of a sphere with grid singularity at the centre, 
\citet{CaiChanDeng2011} proposed a different approach. They use spectral
expansion methods on a spherical grid. To avoid the time step restriction
due to converging grid lines at the centre, they lower the order of the
harmonic expansion at the centre. The equations are recast in a form such
that boundary conditions can easily be applied at the centre. In this way,
the simulation domain can be extended to the full sphere. Anyway, their 
procedure still requires the specification of a boundary condition at the 
centre and applies only to spherical domains.

\citet{MoczVogelsbergerSijackietal2013} implemented a discontinuous Galerkin 
method on arbitrary static and moving Voronoi meshes. In theory, their approach
promises great flexibility and wide applicability. In applications, however,
there are still numerical difficulties present, e.g.\ in the treatment of 
shocks, making the use of the method in astrophysical applications difficult 
at the moment.

In this paper, we present the methods used to extend the applicability of the 
simulation code ANTARES \citep{MuthsamKupkaLoew-Basellietal2010} to more general 
geometries. Until now, ANTARES was exclusively applied to numerical simulations 
of solar and stellar surface convection and stellar interiors in Cartesian geometry 
\citep[e.g.,][]{MuthsamKupkaLoew-Basellietal2010,HappenhoferGrimm-StreleKupkaetal2013}
as well as convection in Cepheids in spherical geometry
\citep[e.g.,][]{MuthsamKupkaMundprechtetal2010,MundprechtMuthsamKupka2013}. For the 
inviscid part of the Navier--Stokes equations, the WENO finite difference scheme is 
employed \citep{ShuOsher1988,Shu2003,Merriman2003}. The WENO scheme is a highly efficient 
shock-capturing scheme which can be implemented at several different orders of 
accuracy. In this paper, we consider the fifth order variant called WENO5. Its 
superiority compared to other high-order schemes was shown, e.g., in 
\citet{MuthsamLoew-BaselliOberscheideretal2007}. Nevertheless,
its applicability is restricted by its specific requirements concerning the grid
geometry \citep{Merriman2003}. 

The technique of curvilinear or mapped grids is widely used in engineering 
\citep{Wesseling2001,FerzigerPeric2002,LeVeque2004}, but until now was only seldomly 
applied in an astrophysical setting \citep{KifonidisMueller2012}. In principle, 
given a suitable mapping function, any problem defined on a general domain can be 
transformed into a problem in a computational space which is equidistant and Cartesian 
and where any standard numerical scheme, the applicability of which often is restricted 
to Cartesian and equidistant grids, can be used. The only requirement is that the 
grid in physical space is structured. 

In the context of curvilinear coordinates, \citet{Calhounetal2008} presented several
functions mapping spherical domains to the Cartesian computational domain. These
functions are not strongly differentiable and were mainly applied in engineering 
applications until now.

In this paper, we will show how well the WENO finite difference and finite
volume scheme performs on smooth and non-smooth grids for flows in the 
intermediate flow regime of $0.1 \leq {\rm Ma} \leq 1$ typical for many problems
in stellar astrophysics. We will call a grid (non)smooth if the associated mapping 
function is (non)smooth.

\citet{Shu2003} applied the WENO finite difference scheme in a straightforward
way to smooth grids.  In all numerical examples in the mentioned paper, the Mach 
number ${\rm Ma} = \frac{\abs{u}}{v_{\rm snd}}$ was higher than $1$.
The behaviour of the method in the low Mach number limit on Cartesian grids
was investigated in~\citet{HappenhoferGrimm-StreleKupkaetal2013}. They showed
that the WENO5 finite difference scheme does not perform well for Mach numbers 
smaller than $0.1$ even on Cartesian grids.

Most of the findings of this paper are not restricted to our specific code,
but apply to any finite difference or finite volume code. From the numerical 
experiments in Section~\ref{sec-results}, we conclude in which situations and 
in which numerical setup the mapped grid technique gives reliable results. 
Thereby, we concentrate on the WENO algorithm and on astrophysical simulations. 
We demonstrate the usefulness and applicability of the mapping functions for 
a sphere from~\citet{Calhounetal2008} in this setting.

\subsection{WENO Finite Difference and Finite Volume Formulation}
\label{sub-weno}

The Euler equations are a system of partial differential equations.
In two spatial dimensions and in a Cartesian coordinate system, 
their differential form is

\begin{equation}
  \frac{\partial}{\partial t}{\bf Q} 
        + \frac{\partial}{\partial x}{\bf F} 
        + \frac{\partial}{\partial y}{\bf G} = 0,\label{eulerdiff2d}
\end{equation}

\noindent with the state vector ${\bf Q}$ and the flux functions ${\bf F}$ and 
${\bf G}$ given by

\begin{equation}
  {\bf Q} = \left( \begin{array}{c}
                      \rho   \\
                      \rho u \\
                      \rho v \\
                      E
                   \end{array} \right),
  {\bf F} \left( {\bf Q} \right) = \left( \begin{array}{c}
                                                 \rho u       \\
                                                 \rho u^2 + p \\
                                                 \rho u v     \\
                                                 ( p + E ) u
                                          \end{array} \right),
  {\bf G} \left( {\bf Q} \right) = \left( \begin{array}{c}
                                                 \rho v       \\
                                                 \rho v u     \\
                                                 \rho v^2 + p \\
                                                 ( p + E ) v
                                          \end{array} \right),
\end{equation}

\noindent where the pressure $p=p(\rho,e)$ is given by an equation of 
state and $e = E - \frac{u^2 + v^2 + w^2}{2 \rho}$ is the internal energy. 
In the following, we will write ${\bf u} := \left( u, v \right)^T$ for 
the velocity vector.

We discretise the Euler equations as described in~\ref{app-disc} and 
in \citet{Merriman2003} for the case of an equidistant Cartesian grid. 
In the finite difference case, the fluxes are reconstructed directly, 
whereas in the finite volume case, the conservative variables are reconstructed.
The procedure is described in pseudo code in Algorithms~\ref{alg-fd-euler} 
and~\ref{alg-fv-euler}.

\begin{algorithm}[ht]
\caption{Finite difference scheme for the two-dimensional 
         Euler equations.}\label{alg-fd-euler}
\begin{algorithmic}[1]
  \STATE ${\bf Q}_{i,j}$ is given as point value at the cell centre.
  \STATE $\mathrm{A}({\bf f})_{i,j} = {\bf F}_{i,j}$, 
         $\mathrm{A}({\bf g})_{i,j} = {\bf G}_{i,j}$
  \STATE ${\bf f}_{i \pm \frac{1}{2},j} 
         = \mathrm{R}_{x} \left( {\bf F}_{i,j} \right)$,
         ${\bf g}_{i,j \pm \frac{1}{2}} 
         = \mathrm{R}_{y} \left( {\bf G}_{i,j} \right)$
  \STATE $\frac{\partial {\bf Q}_{i,j}}{\partial t} = 
         - \frac{1}{\delta x} \left( {\bf f}_{i+\frac{1}{2},j} 
                                   - {\bf f}_{i-\frac{1}{2},j} \right)
         - \frac{1}{\delta y} \left( {\bf g}_{i,j+\frac{1}{2}} 
                                   - {\bf g}_{i,j-\frac{1}{2}} \right)$
\end{algorithmic}
\end{algorithm}

\begin{algorithm}[ht]
\caption{Finite volume scheme for the two-dimensional
         Euler equations.}\label{alg-fv-euler}
\begin{algorithmic}[1]
  \STATE ${\bf Q}_{i,j} = {\bf \overline Q}_{i,j}$ is given as cell average.
  \STATE ${\bf Q}_{i \pm \frac{1}{2},j} 
         = \mathrm{R}_{x} \left( {\bf \overline Q}_{i,j} \right)$,
         ${\bf Q}_{i,j \pm \frac{1}{2}} 
         = \mathrm{R}_{y} \left( {\bf \overline Q}_{i,j} \right)$
  \STATE ${\bf F}_{i \pm \frac{1}{2},j} 
         = {\bf F} \left( {\bf Q}_{i \pm \frac{1}{2},j} \right)$,
         ${\bf G}_{i,j \pm \frac{1}{2}} 
         = {\bf G} \left( {\bf Q}_{i,j \pm \frac{1}{2}} \right)$
  \STATE $\frac{\partial {\bf \overline Q}_{i,j}}{\partial t} = 
         - \frac{1}{\delta x} \left( {\bf F}_{i+\frac{1}{2},j} 
                                   - {\bf F}_{i-\frac{1}{2},j} \right)
         - \frac{1}{\delta y} \left( {\bf G}_{i,j+\frac{1}{2}} 
                                   - {\bf G}_{i,j-\frac{1}{2}} \right)$
\end{algorithmic}
\end{algorithm}

Both algorithms need the specification of a reconstruction operator. A 
reconstruction operator calculates the point value of a function given
its cell averages. The WENO reconstruction operator is described, e.g.,
in \citet{Shu2003} and~\ref{app-rec5th}. In the derivation of the finite 
difference scheme, we required the grid to be equidistant. The fluxes in 
the finite volume scheme are second-order approximations to the analytical 
fluxes which are line integrals over the cell boundary. Details can be found 
in~\ref{app-disc}.

\section{Mapped Grids}
\label{sec-mapped}

Numerical schemes which are designed for Cartesian, equidistant grids can be generalised 
to more complicated domains with the technique of mapped grids. There, a mapping function 

\begin{equation}
  M:\ \left[ -1, 1 \right]^2 \to \Omega,\ M(\xi,\eta) = (x, y)^T,
\end{equation}

\noindent is defined which maps the Cartesian and equidistant computational space into the 
physical space. The information about the geometry of the physical space is then 
contained in the transformed partial differential equations. The Euler equations in strong 
conservation form in physical space~\eqref{eulerdiff2d} are transformed into strong 
conservation form in computational space. In two dimensions, they take the form

\begin{subequations}
\begin{equation}
  \frac{\partial}{\partial t}J^{-1}{\bf Q} + \frac{\partial}{\partial \xi }{\bf \hat{F}} 
                                           + \frac{\partial}{\partial \eta}{\bf \hat{G}} = 0
\end{equation}

\noindent with

\begin{align}
  {\bf \hat{F}} = &   \frac{\partial y}{\partial \eta} {\bf F} - \frac{\partial x}{\partial \eta} {\bf G}, \\
  {\bf \hat{G}} = & - \frac{\partial y}{\partial \xi}  {\bf F} + \frac{\partial x}{\partial \xi}  {\bf G},
\end{align}\label{eulercurvi2d}
\end{subequations}

\noindent where $J^{-1}$ is the determinant of the inverse Jacobian of the mapping 
function $M$ \citep[see, e.g.,][]{KifonidisMueller2012}. $\xi$ and $\eta$ are the 
computational variables defined by the mapping function $M$.

We present two derivations of the strong conservation form of the two-dimensional Euler 
equations in computational space~\eqref{eulercurvi2d} which differ in their differentiability 
assumptions concerning the mapping function $M$. The classical first approach assuming 
strong differentiability of the mapping function can be found in~\ref{app-strong}.
A more general derivation is sketched out in the following section.

In applications, the mapping function $M$ which maps the computational domain into the 
physical domain $\Omega$ is unknown or does not possess an analytical form. We therefore seek 
for a derivation of the transformed equations~\eqref{eulercurvi2d} which does not require 
any differentiability of the mapping function $M$. Following the description in \citet{Wesseling2001}, 
we assume that a structured set of nodes $(x_{i \pm \frac{1}{2},j \pm \frac{1}{2}},
y_{i \pm \frac{1}{2},j \pm \frac{1}{2}})$, $1 \leq i \leq n$, $1 \leq j \leq m$, is given \citep[e.g.,
by an external grid generation program, see][]{ThompsonWarsiMastin1985}, instead of an analytical 
expression for $M$. $\Omega$ is the smallest region $R \subset \R^2$ such that 
$(x_{i \pm \frac{1}{2},j \pm \frac{1}{2}},y_{i \pm \frac{1}{2},j \pm \frac{1}{2}}) \in \overline{R}\ 
\forall i,j$, where $\overline{R}$ is the closure of $R$. Then we define the discrete mapping function 
$\tilde{M}$ point-wisely by

\begin{equation}
  \tilde{M}(\xi_{i-\frac{1}{2}},\eta_{j-\frac{1}{2}}) 
  = (x_{i-\frac{1}{2},j-\frac{1}{2}},y_{i-\frac{1}{2},j-\frac{1}{2}}),
  i = 1, \hdots, n+1, j = 1, \hdots, m+1.
\end{equation}

For $(\xi,\eta) \in C_{i,j}:= \left[  \xi_{i-\frac{1}{2}}, \xi_{i+\frac{1}{2}} \right]
                       \times \left[ \eta_{j-\frac{1}{2}},\eta_{j+\frac{1}{2}} \right]$, we define 
$M(\xi,\eta)$ by bilinear interpolation of the physical coordinates of the edges of the cell. In 
this way, we continue $\tilde{M}$ to $M:\ \left[ -1, 1 \right] \times \left[ -1, 1 \right] \to \Omega$.
$M$ is continuous, linear in each $C_{i,j}$, but not differentiable on the boundary 
of each cell $C_{i,j}$. It follows that $M \not\in C^1(\left[ -1, 1 \right]^2)$, but 
$M \in H^1(\left[ -1, 1 \right]^2)$.

If the mapping function is defined in this way, the image of each cell $C_{i,j}$ is a quadrilateral 
$D_{i,j}$ with edges $\left( x_{i-\frac{1}{2},j-\frac{1}{2}}, y_{i-\frac{1}{2},j-\frac{1}{2}} \right)$,
                     $\left( x_{i-\frac{1}{2},j+\frac{1}{2}}, y_{i-\frac{1}{2},j+\frac{1}{2}} \right)$,
                     $\left( x_{i+\frac{1}{2},j-\frac{1}{2}}, y_{i+\frac{1}{2},j-\frac{1}{2}} \right)$,
                 and $\left( x_{i+\frac{1}{2},j+\frac{1}{2}}, y_{i+\frac{1}{2},j+\frac{1}{2}} \right)$
in physical space. All cell sides are straight lines. We can choose the quadrilateral $D_{i,j}$ as 
the (arbitrary) control volume $W \subset \Omega$ of the integral formulation of the Euler equations

\begin{subequations}
\begin{align}
\frac{\partial}{\partial t} \int_{W} \rho        \,dV = & 
    - \int_{\partial W} {\bf n} \cdot \rho{\bf u}\,dA,  \\
\frac{\partial}{\partial t} \int_{W} \rho {\bf u}\,dV = & 
    - \int_{\partial W} \left[ {\bf n} \cdot \rho {\bf u} \otimes {\bf u} + {\bf n} p \right]\,dA,  \\
\frac{\partial}{\partial t} \int_{W} E           \,dV = & 
    - \int_{\partial W} {\bf n} \cdot \left( p + E \right) {\bf u}\,dA,
\end{align}\label{EulerInt}
\end{subequations}

\noindent where ${\bf n}$ is the unit outward normal on $\partial W$, the boundary of $W$.
Due to Gauss' Theorem \citep[p. 627,][]{Evans2002}, equations~\eqref{EulerInt} are equivalent
to the differential form~\eqref{eulerdiff2d} for sufficiently smooth functions
\citep{ChorinMarsden1993}. But $\partial W = \partial D_{i,j}$ consists of the four straight 
lines 

\begin{subequations}
\begin{align}
  S_{i \pm \frac{1}{2}} & = \left( \begin{array}{c} x_{i \pm \frac{1}{2},j+\frac{1}{2}} \\
                                                    y_{i \pm \frac{1}{2},j+\frac{1}{2}}
                                   \end{array} \right)
                          - \left( \begin{array}{c} x_{i \pm \frac{1}{2},j-\frac{1}{2}} \\
                                                    y_{i \pm \frac{1}{2},j-\frac{1}{2}}
                                   \end{array} \right), \\
  S_{j \pm \frac{1}{2}} & = \left( \begin{array}{c} x_{i+\frac{1}{2},j \pm \frac{1}{2}} \\
                                                    y_{i+\frac{1}{2},j \pm \frac{1}{2}}
                                   \end{array} \right)
                          - \left( \begin{array}{c} x_{i-\frac{1}{2},j \pm \frac{1}{2}} \\
                                                    y_{i-\frac{1}{2},j \pm \frac{1}{2}}
                                   \end{array} \right).
\end{align}
\end{subequations}

Their length is given by 

\begin{subequations}
\begin{align}
  \abs{S_{i \pm \frac{1}{2}}} & = \sqrt{ \left( x_{i \pm \frac{1}{2},j+\frac{1}{2}} 
                                              - x_{i \pm \frac{1}{2},j-\frac{1}{2}} \right)^2
                                       + \left( y_{i \pm \frac{1}{2},j+\frac{1}{2}} 
                                              - y_{i \pm \frac{1}{2},j-\frac{1}{2}} \right)^2 }, \\
  \abs{S_{j \pm \frac{1}{2}}} & = \sqrt{ \left( x_{i+\frac{1}{2},j \pm \frac{1}{2}} 
                                              - x_{i-\frac{1}{2},j \pm \frac{1}{2}} \right)^2
                                       + \left( y_{i+\frac{1}{2},j \pm \frac{1}{2}} 
                                              - y_{i-\frac{1}{2},j \pm \frac{1}{2}} \right)^2 },
\end{align}

\noindent and the normal vectors ${\bf n} = \left( \begin{array}{c} n_1 \\
                                                                    n_2 \end{array} \right)$
in equations~\eqref{EulerInt} are exactly

\begin{align}
  n_1 & =   \left( y_{i \pm \frac{1}{2},j+\frac{1}{2}} - y_{i \pm \frac{1}{2},j-\frac{1}{2}} \right) / \abs{S_{i \pm \frac{1}{2}}}, \\
  n_2 & = - \left( x_{i \pm \frac{1}{2},j+\frac{1}{2}} - x_{i \pm \frac{1}{2},j-\frac{1}{2}} \right) / \abs{S_{i \pm \frac{1}{2}}}
\end{align}

\noindent on $S_{i \pm \frac{1}{2}}$ and

\begin{align}
  n_1 & = - \left( y_{i+\frac{1}{2},j \pm \frac{1}{2}} - y_{i-\frac{1}{2},j \pm \frac{1}{2}} \right) / \abs{S_{j \pm \frac{1}{2}}}, \\
  n_2 & =   \left( x_{i+\frac{1}{2},j \pm \frac{1}{2}} - x_{i-\frac{1}{2},j \pm \frac{1}{2}} \right) / \abs{S_{j \pm \frac{1}{2}}}.
\end{align}
\end{subequations}

\noindent on $S_{j \pm \frac{1}{2}}$. From now on, we will write 
${\bf n} = \left( \begin{array}{c} n_1 \\
                                   n_2 \end{array} \right)$ 
for the normal vector on $S_{i \pm \frac{1}{2}}$ and 
${\bf m} = \left( \begin{array}{c} m_1 \\
                                   m_2 \end{array} \right)$ 
for the normal vector on $S_{j \pm \frac{1}{2}}$.
The surface area of the quadrilateral $D_{i,j}$ is 

\begin{align*}
  \abs{D_{i,j}} = & \frac{1}{2} \Big| ( y_{i-\frac{1}{2},j-\frac{1}{2}}-y_{i+\frac{1}{2},j+\frac{1}{2}} )
                                      ( x_{i-\frac{1}{2},j+\frac{1}{2}}-x_{i+\frac{1}{2},j-\frac{1}{2}} ) \\
    &                               + ( y_{i+\frac{1}{2},j-\frac{1}{2}}-y_{i-\frac{1}{2},j+\frac{1}{2}} )
                                      ( x_{i-\frac{1}{2},j-\frac{1}{2}}-x_{i+\frac{1}{2},j+\frac{1}{2}} ) \Big|.
\end{align*}

Since $\abs{D_{i,j}} > 0$, we define the cell average of the state vector 
${\bf Q}_{i,j}$ in the cell $(i,j)$ by

\begin{equation}
  \overline{\bf Q}_{i,j} = \abs{D_{i,j}}^{-1} \int_{D_{i,j}} {\bf Q}\,dV.
\end{equation}

With $U := n_1 u + n_2 v$, $V := m_1 u + m_2 v$, we can write equations~\eqref{EulerInt} as

\begin{equation}
    \frac{\partial}{\partial t} \overline{\bf Q}_{i,j} =   
    - \left( \int_{S_{i+\frac{1}{2}}} {\bf \hat{F}} \,dS
           - \int_{S_{i-\frac{1}{2}}} {\bf \hat{F}} \,dS
           + \int_{S_{j+\frac{1}{2}}} {\bf \hat{G}} \,dS
           - \int_{S_{j-\frac{1}{2}}} {\bf \hat{G}} \,dS \right),\label{eq-curvdisc1}
\end{equation}

\noindent where

\begin{equation}
  {\bf Q} = \left( \begin{array}{c}
                      \rho   \\
                      \rho u \\
                      \rho v \\
                      E
                   \end{array} \right),
  {\bf \hat{F}} = \left( \begin{array}{c}
                              \rho U           \\
                              \rho U u + n_1 p \\
                              \rho U v + n_2 p \\
                              ( p + E ) U
                         \end{array} \right),
  {\bf \hat{G}} = \left( \begin{array}{c}
                              \rho V           \\
                              \rho V u + m_1 p \\
                              \rho V u + m_2 p \\
                              ( p + E ) V
                         \end{array} \right).
\end{equation}

Evaluating the line integrals in~\eqref{eq-curvdisc1} with the midpoint rule, we get

\begin{equation}
    \frac{\partial}{\partial t} \abs{D_{i,j}} \overline{\bf Q}_{i,j} =   
    - \left( {\bf \hat{F}}_{i+\frac{1}{2},j} - {\bf \hat{F}}_{i-\frac{1}{2},j}
           + {\bf \hat{G}}_{i,j+\frac{1}{2}} - {\bf \hat{G}}_{i,j-\frac{1}{2}} \right).\label{eq-curvdisc2}
\end{equation}

In the computational space, all standard numerical methods can be used to calculate 
the value of the numerical flux functions ${\bf \hat{F}}$ and ${\bf \hat{G}}$ 
since the computational space is equidistant and Cartesian. In particular, both the 
finite difference and the finite volume WENO scheme as described in 
Algorithms~\ref{alg-fd-euler} and~\ref{alg-fv-euler} can be applied to 
solve~\eqref{eq-curvdisc2}. The specific form of the algorithms for curvilinear 
coordinates can be found in Algorithms~\ref{alg-fd-curvi} and~\ref{alg-fv-curvi}.

\begin{algorithm}[ht]
\caption{Finite difference scheme for curvilinear coordinates.}\label{alg-fd-curvi}
\begin{algorithmic}[1]
  \STATE ${\bf Q}_{i,j}$ is given as point value at the cell centre.
  \STATE $\mathrm{A}({\bf \hat f})_{i,j} = {\bf \hat F}_{i,j} 
         = {n_1}|_{i,j} {\bf F}_{i,j} + {n_2}|_{i,j} {\bf G}_{i,j}$, \\
         $\mathrm{A}({\bf \hat g})_{i,j} = {\bf \hat G}_{i,j} 
         = {m_1}|_{i,j} {\bf F}_{i,j} + {m_2}|_{i,j} {\bf G}_{i,j}$
  \STATE ${\bf \hat f}_{i \pm \frac{1}{2},j} 
         = \mathrm{R}_{\xi } \left( {\bf \hat F}_{i,j} \right)$,
         ${\bf \hat g}_{i,j \pm \frac{1}{2}} 
         = \mathrm{R}_{\eta} \left( {\bf \hat G}_{i,j} \right)$
  \STATE $\frac{\partial {\bf Q}_{i,j}}{\partial t} = 
         - \frac{1}{\delta \xi } \left( {\bf \hat f}_{i+\frac{1}{2},j} 
                                      - {\bf \hat f}_{i-\frac{1}{2},j} \right)
         - \frac{1}{\delta \eta} \left( {\bf \hat g}_{i,j+\frac{1}{2}} 
                                      - {\bf \hat g}_{i,j-\frac{1}{2}} \right)$
\end{algorithmic}
\end{algorithm}

\begin{algorithm}[ht]
\caption{Finite volume scheme for curvilinear coordinates.}\label{alg-fv-curvi}
\begin{algorithmic}[1]
  \STATE ${\bf Q}_{i,j} = {\bf \overline Q}_{i,j}$ is given as cell average.
  \STATE ${\bf Q}_{i \pm \frac{1}{2},j} 
         = \mathrm{R}_{\xi } \left( {\bf \overline Q}_{i,j} \right)$,
         ${\bf Q}_{i,j \pm \frac{1}{2}} 
         = \mathrm{R}_{\eta} \left( {\bf \overline Q}_{i,j} \right)$

  \STATE ${\bf \hat F}_{i \pm \frac{1}{2},j} 
         = {n_1}|_{i \pm \frac{1}{2},j} {\bf F} \left( {\bf Q}_{i \pm \frac{1}{2},j} \right) 
         + {n_2}|_{i \pm \frac{1}{2},j} {\bf G} \left( {\bf Q}_{i \pm \frac{1}{2},j} \right)$, \\
         ${\bf \hat G}_{i,j \pm \frac{1}{2}} 
         = {m_1}|_{i,j \pm \frac{1}{2}} {\bf F} \left( {\bf Q}_{i,j \pm \frac{1}{2}} \right) 
         + {m_2}|_{i,j \pm \frac{1}{2}} {\bf G} \left( {\bf Q}_{i,j \pm \frac{1}{2}} \right)$
  \STATE $\frac{\partial {\bf \overline Q}_{i,j}}{\partial t} = 
         - \frac{1}{\delta \xi } \left( {\bf \hat F}_{i+\frac{1}{2},j} 
                                      - {\bf \hat F}_{i-\frac{1}{2},j} \right)
         - \frac{1}{\delta \eta} \left( {\bf \hat G}_{i,j+\frac{1}{2}} 
                                      - {\bf \hat G}_{i,j-\frac{1}{2}} \right)$
\end{algorithmic}
\end{algorithm}

The advantage of the weak derivation, besides that it does not require $M$ to be 
differentiable, is that one gets formulae for all metric terms occurring in the 
transformation process. The Jacobian of the transformation is given by the area of 
the corresponding quadrilateral and therefore is positive by definition. The mapping 
function is not required to fulfil any smoothness properties. We therefore prefer to 
define our mapping function in this way.

We note that similar formulae exist for the three-dimensional case. The precise 
formulation can be found, e.g., in \citet{VisbalGaitonde2002}.

\subsection{Grids for Spherical Domains}

There is no strongly differentiable function (a diffeomorphism) from the (unit) sphere 
to the (unit) square since the square is not a submanifold of $\R^2$. Therefore, if we 
want to use the mapped grid technique to perform simulations on spherical domains, we 
have to rely on mapping functions which are only weakly differentiable.

For the mapped grid technique, \citet{Calhounetal2008} gave some examples of mapping 
functions which map a circular domain to $\left[ -1, 1 \right]^2$ and vice versa. In 
this paper, we want to investigate how the mapping functions $M_1$ and $M_2$ from 
\citet{Calhounetal2008} defined by

\begin{subequations}
\begin{align}
  M_1:\ \left[ -1, 1 \right]^2 \to \Omega,\  
              & x = R \cdot \frac{\max \left( \left|\xi\right|,\left|\eta\right| \right) \xi }{\sqrt{ \xi^2 + \eta^2}}, \\
              & y = R \cdot \frac{\max \left( \left|\xi\right|,\left|\eta\right| \right) \eta}{\sqrt{ \xi^2 + \eta^2}}, \\
  M_2:\ \left[ -1, 1 \right]^2 \to \Omega,\  
              & w = \max \left( \left|\xi\right|,\left|\eta\right| \right)^2, \\
              & x = w \cdot x_{M_1} + (1 - w) \cdot \frac{R \xi }{\sqrt{2}},   \\
              & y = w \cdot y_{M_1} + (1 - w) \cdot \frac{R \eta}{\sqrt{2}},
\end{align}\label{eq-calhoun}
\end{subequations}

\noindent where $x_{M_1}$ and $y_{M_1}$ are the physical coordinates defined by 
$M_1$, and

\begin{equation}
  \Omega = \{ \left( x, y \right) \in \R^2: \sqrt{x^2+y^2} \leq R \},
\end{equation}

\noindent perform in numerical simulations when WENO schemes are employed.

It is obvious to see that these functions are only weakly differentiable. 
Therefore, they should be applied only in the context of the methods developed 
in Section~\ref{sec-mapped}.

\begin{figure}
\vspace{0pt}
\centering
\includegraphics[width=0.5\textwidth]{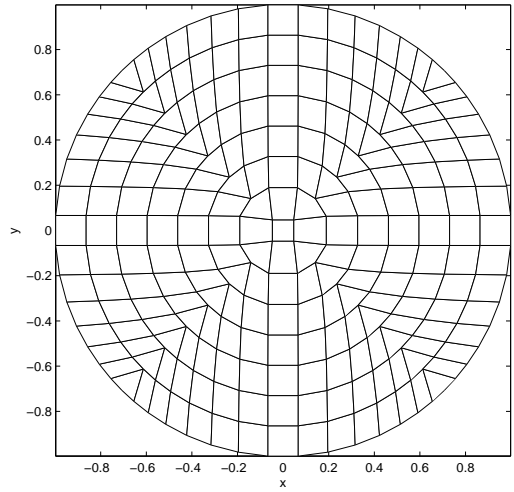}
\caption{Grid defined by function $M_1$ from \citet{Calhounetal2008}.}\label{fig-map1}
\end{figure}

\begin{figure}
 \vspace{0pt}
\centering
\includegraphics[width=0.5\textwidth]{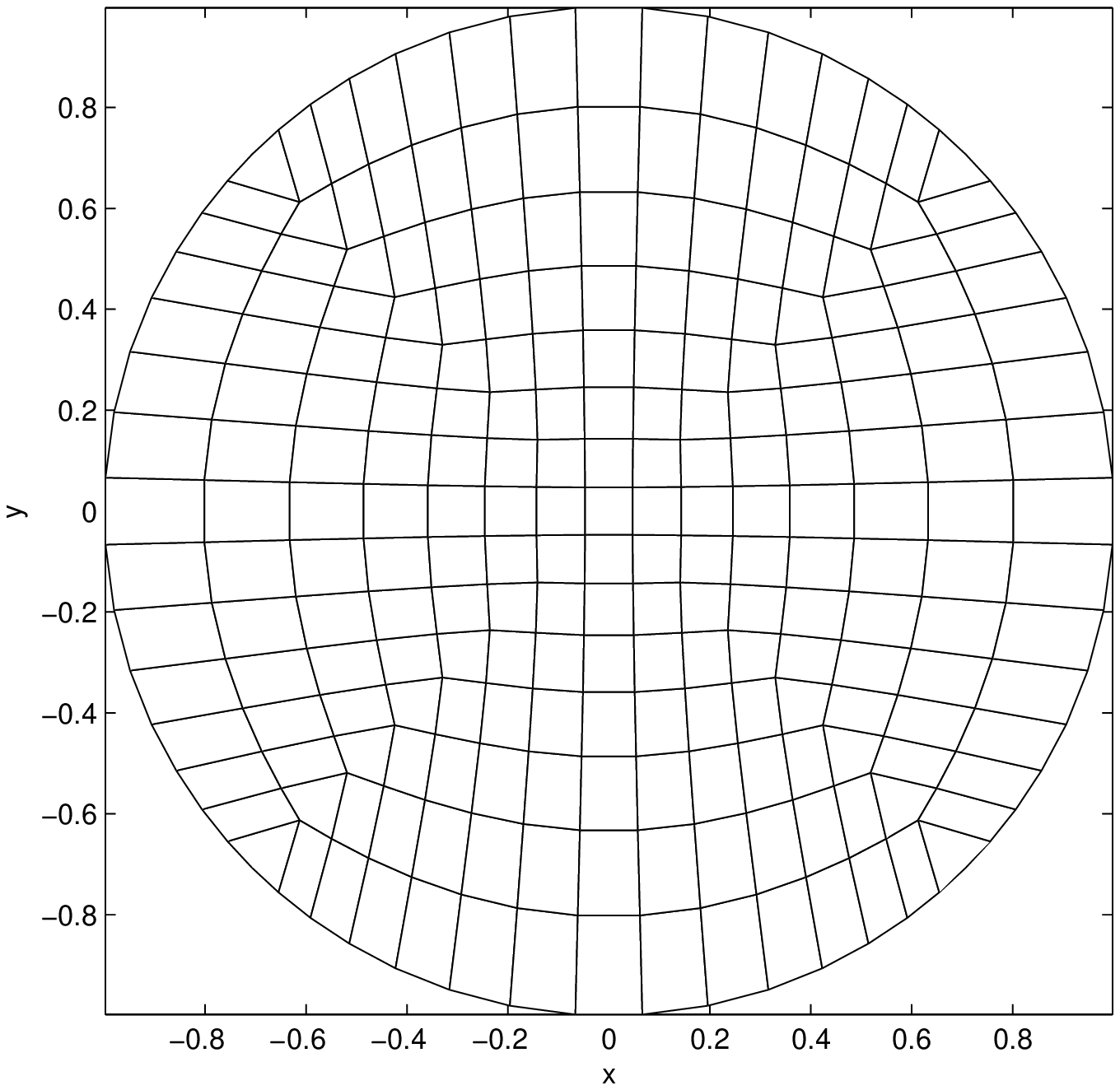}
\caption{Grid defined by function $M_2$ from \citet{Calhounetal2008}.}\label{fig-map2}
\end{figure}

\subsection{Freestream Problem}

\citet{NonomuraIizukaFujii2010} emphasize the importance of the following, 
seemingly simple, test problem.

\begin{problem}[Freestream preservation for the Euler equations]\label{freestreameuler}
  Given the initial conditions

  \begin{equation}
     \left( \rho, \rho u, \rho v, E \right) = \left( \rho_0, 0, 0, e_0 \right),   
  \end{equation}

  \noindent with constant density $\rho_0$ and internal energy $e_0$, the numerical 
  solution of the transformed system~\eqref{eulercurvi2d} should stay close to the 
  analytical solution

  \begin{equation}
    \left( \rho, \rho u, \rho v, E \right) = \left( \rho_0, 0, 0, e_0 \right)
  \end{equation}
 
  \noindent for all times. $p_0=p(\rho_0,e_0)$ is the pressure given by the 
  equation of state.
\end{problem}

Plugging these initial conditions into the transformed Euler 
equations~\eqref{eulercurvi2d}, we get

\begin{equation}
  0 = p_0 \left( \pad[n_1]{\xi} + \pad[m_1]{\eta} \right),\ 
  0 = p_0 \left( \pad[n_2]{\xi} + \pad[m_2]{\eta} \right),
\end{equation}

\noindent since $\pad{t} \left( \rho u \right) = \pad{t} \left( \rho v \right) = 0$. This 
condition must be fulfilled numerically in order to prevent numerical errors. Since

\begin{equation}
  n_1 =   \pad[y]{\eta}, n_2 = - \pad[x]{\eta}, 
  m_1 = - \pad[y]{\xi }, m_2 =   \pad[x]{\xi }, 
\end{equation}

\noindent this is equivalent to the requirement that the second derivatives
of the mapping function $M$ are symmetric. Every function which is 
twice (weakly) differentiable fulfils this property.

Next, we investigate if the freestream is preserved by the finite difference 
and the finite volume discretisation of the Euler equations. The WENO finite 
difference scheme as summarised in Algorithm~\ref{alg-fd-curvi} corresponds 
to the scheme \texttt{WENO-G} in \citet{NonomuraIizukaFujii2010}. In the cited 
reference, they describe precisely why the finite difference scheme does not 
fulfil the freestream preservation property. The fluxes $\hat{\bf F}$ and 
$\hat{\bf G}$ are reconstructed directly from their value at the cell centre, 
including the metric terms evaluated at the cell centre. These fluxes are not
constant for the freestream initial conditions. The reconstructed fluxes at the
cell boundary will not be constant, too, and be different on any cell boundary. 
Therefore, they will not cancel out, and steadily, a numerical error is 
introduced.

We note that \citet{NonomuraIizukaFujii2010} introduced the \texttt{WENO-C} 
scheme as a different WENO finite difference scheme which fulfils the 
freestream property by calculating $\pad[{\bf Q}]{t}$ via

\begin{equation}
  \pad[{\bf Q}]{t} = \pad[\xi ]{x} \pad[{\bf F}]{\xi} + \pad[\eta]{x} \pad[{\bf F}]{\eta}
                   + \pad[\xi ]{y} \pad[{\bf G}]{\xi} + \pad[\eta]{y} \pad[{\bf G}]{\eta}.
\end{equation}

We did not consider this scheme in our paper since it is not conservative and its 
computational costs are three times higher than
\texttt{WENO-G} in three dimensions, making the scheme useless for our purposes.

On the contrary, for the WENO finite volume scheme as summarised in 
Algorithm~\ref{alg-fv-curvi}, the state vector ${\bf Q}$ is reconstructed 
at the cell boundaries from its cell averages. Therefore, for 
Problem~\ref{freestreameuler} the reconstruction process will yield constant 
approximations for the state vector at the cell interface, i.e.

\begin{equation}
  {\bf Q}_{i \pm \frac{1}{2},j} = {\bf Q}_{i,j \pm \frac{1}{2}} = \left( \rho_0, 0, 0, e_0 \right)^T.
\end{equation}
 
In the update step~\eqref{eq-curvdisc2}, only the metric terms will be 
non-constant. In precise terms, the conditions 

\begin{subequations}
\begin{align}
  \frac{n_1|_{i+\frac{1}{2},j}-n_1|_{i-\frac{1}{2},j}}{\delta \xi} 
                       + \frac{m_1|_{i,j+\frac{1}{2}}-m_1|_{i,j-\frac{1}{2}}}{\delta \eta} = 0, \\
  \frac{n_2|_{i+\frac{1}{2},j}-n_2|_{i-\frac{1}{2},j}}{\delta \xi} 
                       + \frac{m_2|_{i,j+\frac{1}{2}}-m_2|_{i,j-\frac{1}{2}}}{\delta \eta} = 0,
\end{align}\label{metcond}
\end{subequations}

\noindent are equivalent to preserving the freestream. If the metric terms are calculated, 
as described in Section~\ref{sec-mapped}, by

\begin{subequations}
\begin{align}
  n_1|_{i \pm \frac{1}{2},j} = & \pad[y]{\eta}|_{i \pm \frac{1}{2},j} 
  =   \frac{y_{i \pm \frac{1}{2},j+\frac{1}{2}}-y_{i \pm \frac{1}{2},j-\frac{1}{2}}}{\delta \eta}, \\
  n_2|_{i \pm \frac{1}{2},j} = & - \pad[x]{\eta}|_{i \pm \frac{1}{2},j} 
  = - \frac{x_{i \pm \frac{1}{2},j+\frac{1}{2}}-x_{i \pm \frac{1}{2},j-\frac{1}{2}}}{\delta \eta}, \\
  m_1|_{i,j \pm \frac{1}{2}} = & - \pad[y]{\xi}|_{i,j \pm \frac{1}{2}} 
  = - \frac{y_{i+\frac{1}{2},j \pm \frac{1}{2}}-y_{i-\frac{1}{2},j \pm \frac{1}{2}}}{\delta \xi }, \\
  m_2|_{i,j \pm \frac{1}{2}} = & \pad[x]{\xi }|_{i,j \pm \frac{1}{2}} 
  =   \frac{x_{i+\frac{1}{2},j \pm \frac{1}{2}}-x_{i-\frac{1}{2},j \pm \frac{1}{2}}}{\delta \xi },
\end{align}\label{eq-metric2d}
\end{subequations}

\noindent conditions~\eqref{metcond} are fulfilled exactly and the freestream 
will be preserved numerically.

We note that the freestream is never preserved in general if analytical 
expressions for the metric terms are used (if available).

We remark that even though conditions~\eqref{eq-metric2d} look like second-order 
approximations, they are rather analytical requirements which must be fulfilled 
by the discretisation of the metric terms in order to preserve the freestream. 
They are a consequence of the conservative discretisation of the derivatives in 
equations~\eqref{eulerdiff2d}. As described in~\ref{app-disc}, the discrete
formulations of the Euler equations as in~\eqref{startFV} and~\eqref{startFD} 
are analytically equivalent to~\eqref{eulerdiff2d}.

For trivial mapping functions such as $M:\ \left[ -1, 1 \right]^2 \to
\left[ -1, 1 \right]^2$, $M(\xi,\eta) = (\xi,\eta)^T$ with
$\pad[y]{\eta} = \pad[x]{\xi} = {\rm const}$, $\pad[y]{\xi} = \pad[x]{\eta} = 0$, 
freestream preservation is of course possible even for the finite difference 
scheme \texttt{WENO-G}.

We conclude that the mapped grid technique should be used only with the WENO 
finite volume scheme and with the metric terms calculated by~\eqref{metcond}. 
Non-preservation of the freestream leads to inacceptable errors, as we will 
demonstrate in the following section. The WENO finite difference scheme cannot 
preserve the freestream and be conservative at the same time.

\section{Numerical Results}
\label{sec-results}

In the following, we illustrate the theoretical results from the preceeding sections by numerical
simulations. Besides the mapping functions $M_1$ and $M_2$ defined in~\eqref{eq-calhoun}, we use 
the mapping functions

\begin{align}
  M_3:\ \left[ -1, 1 \right]^2 \to \Omega,\  
  \begin{split}
              & x = -R + 2 R \cdot \left( \xi  + 0.1 \sin \left( 2 \pi \xi \right) \sin \left( 2 \pi \eta \right) \right), \\
              & y = -R + 2 R \cdot \left( \eta + 0.1 \sin \left( 2 \pi \xi \right) \sin \left( 2 \pi \eta \right) \right), \\
  \end{split} \\
  M_4:\ \left[ -1, 1 \right]^2 \to \Omega,\  
  \begin{split}
              & x = -R + 2 R \cdot \xi , \\
              & y = -R + 2 R \cdot \eta,
  \end{split}
\end{align}

\noindent where $\Omega = \left[ -R, R \right]^2$. $M_3$ was used in 
\citet{ColellaDorrHittingerMartin2011} to test the order of accuracy of their scheme 
since it is a smooth function, whereas $M_4$ gives a Cartesian grid. We will call 
a grid ``smooth'' if the mapping function defining this grid is smooth.

\begin{figure}
\vspace{0pt}
\centering
\includegraphics[width=0.5\textwidth]{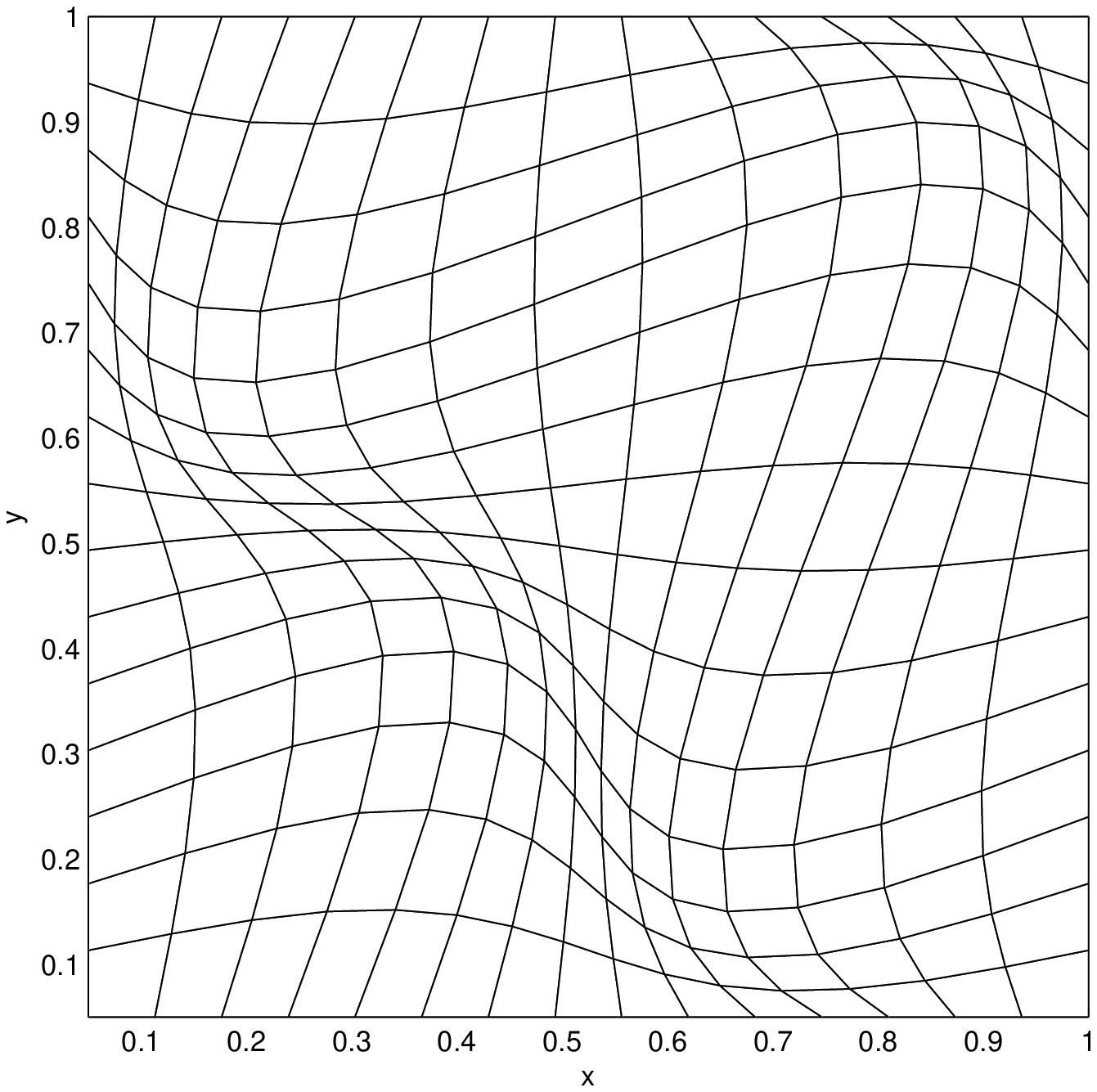}
\caption{Grid defined by function $M_3$ from \citet{ColellaDorrHittingerMartin2011}.}\label{fig-smooth}
\end{figure}

All simulations are performed with the code ANTARES 
\citep{MuthsamKupkaLoew-Basellietal2010} using explicit time integration 
schemes and Marquina flux splitting \citep{DonatMarquina1996}. If not stated 
otherwise, the Runge--Kutta scheme employed in the simulations is SSP RK(3,2), 
a second-order Runge--Kutta scheme with three stages \citep{Kraaijevanger1991,
KupkaHappenhoferHiguerasKoch2012}. 
In all simulations, the ideal gas equation is used, and the Courant number 
is fixed to $0.1$. The WENO finite difference scheme corresponds to the 
method \texttt{WENO-G} in \citet{NonomuraIizukaFujii2010}.

\subsection{Gresho Vortex}

The specific setup of the Gresho Vortex used in this paragraph is described in 
\citet{HappenhoferGrimm-StreleKupkaetal2013} and \citet{Miczek2013}. We repeat the 
definition here for convenience.

\begin{subequations}
\begin{align}
  \rho     = & \,1, \\
  p_0      = & \,\frac{\rho}{\gamma\,{\rm Ma}_{\rm ref}^2}, \\
  u_{\phi} = & \begin{cases}
                 5r,   & 0   \leq r < 0.2, \\
                 2-5r, & 0.2 \leq r < 0.4, \\
                 0,    & 0.4 \leq r,
               \end{cases}, \\
  p        = & \begin{cases}
                 p_0 + \frac{25}{2} r^2,                              & 0   \leq r < 0.2, \\
                 p_0 + \frac{25}{2} r^2 + 4 ( 1-5r-\ln 0.2 + \ln r ), & 0.2 \leq r < 0.4, \\
                 p_0 - 2 + 4 \ln 2,                                   & 0.4 \leq r.
               \end{cases}
\end{align}
\end{subequations}

$r = \sqrt{x^2+y^2}$ is the distance from the origin and $u_{\phi}$ the angular velocity
in terms of the polar angle $\phi = \texttt{atan2}(y,x)$. Note that the difference to the 
setup as it is described in \citet{LiskaWendroff2003} is the introduction of a reference 
Mach number ${\rm Ma}_{\rm ref}$. The pressure $p$ is scaled such that the reference Mach 
number is the maximum Mach number of the resulting flow.

\begin{figure}
\vspace{0pt}
\centering
\includegraphics[width=1.0\textwidth]{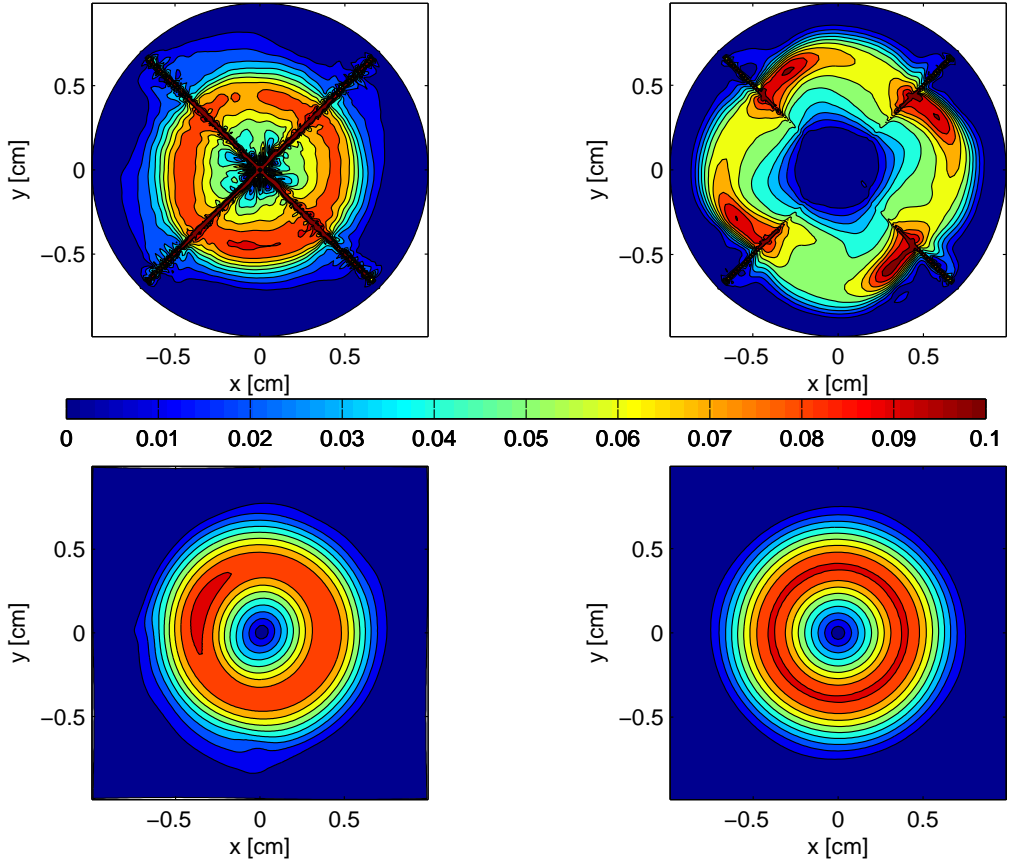}
\caption{Mach number in the Gresho vortex test after $2\,{\rm s}$ with the WENO finite difference 
         scheme. The initial Mach number was $0.1$. In all figures, the results of the 
         simulation with $M_1$ are shown in the top left panel, with $M_2$ in the top right, 
         with $M_3$ in the bottom left and with $M_4$ in the bottom right panel.}
         \label{fig-GreshoMa01-FD}
\end{figure}

\begin{figure}
\vspace{0pt}
\centering
\includegraphics[width=1.0\textwidth]{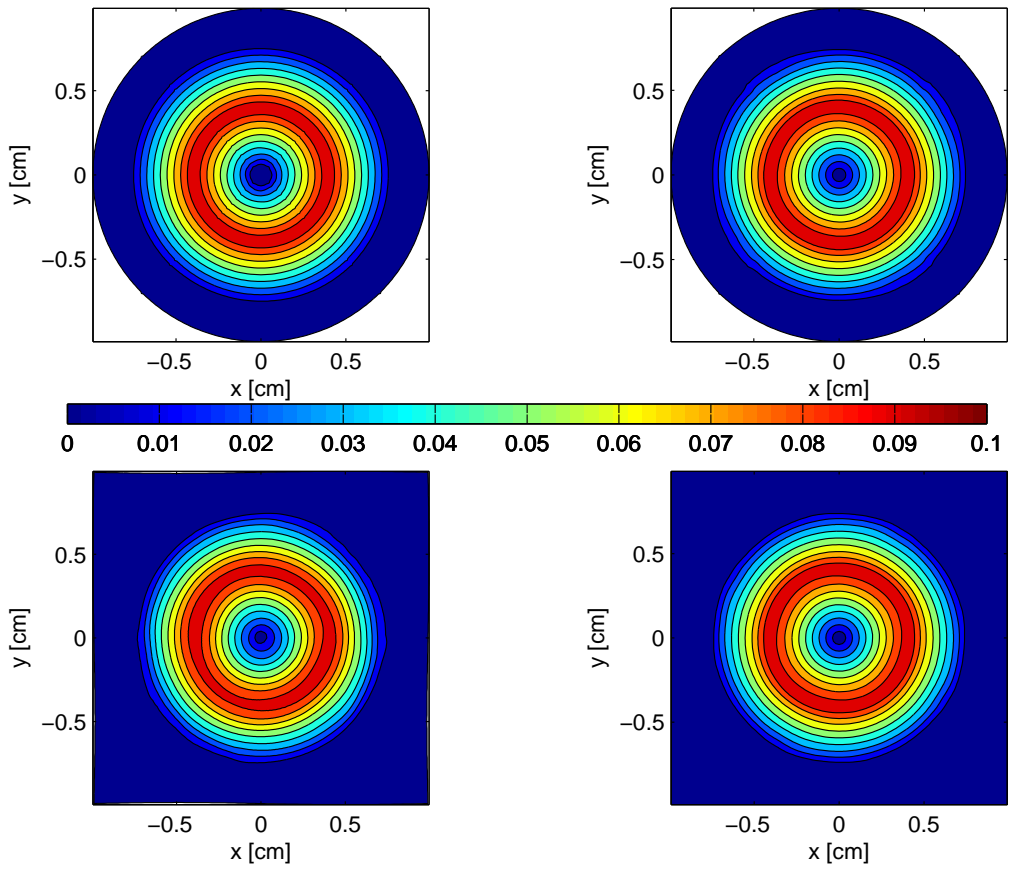}
\caption{Mach number in the Gresho vortex test after $2\,{\rm s}$ with the WENO finite volume 
         scheme. The initial Mach number was $0.1$.}
         \label{fig-GreshoMa01-FV}
\end{figure}

We performed a simulation with ${\rm Ma}_{\rm ref}=0.1$ and $\gamma = \frac{5}{3}$ over a 
time interval of $2\,{\rm s}$. The size of the domain is $1\,{\rm cm}$ in each direction, 
and we use $100 \times 100$ grid points. The analytical solution is pure angular rotation of 
the vortex. The results on the four grids defined by the mapping functions $M_1$, $M_2$, 
$M_3$ and $M_4$ are shown in Figure~\ref{fig-GreshoMa01-FD} for the finite difference 
scheme and in Figure~\ref{fig-GreshoMa01-FV} for the finite volume scheme.

With the non-smooth mapping functions $M_1$ and $M_2$, the results with the finite difference 
scheme are catastrophic, whereas with the finite volume scheme, the difference to the solution 
on the Cartesian grid is small. It is obvious that the problems come from the grid points where 
the mapping functions are not smooth. But even with the smooth mapping function $M_3$, the 
symmetry of the vortex is destroyed with the finite difference scheme, whereas there is no 
visible difference to the Cartesian solution with the finite volume scheme. 

\begin{figure}
\vspace{0pt}
\centering
\includegraphics[width=1.0\textwidth]{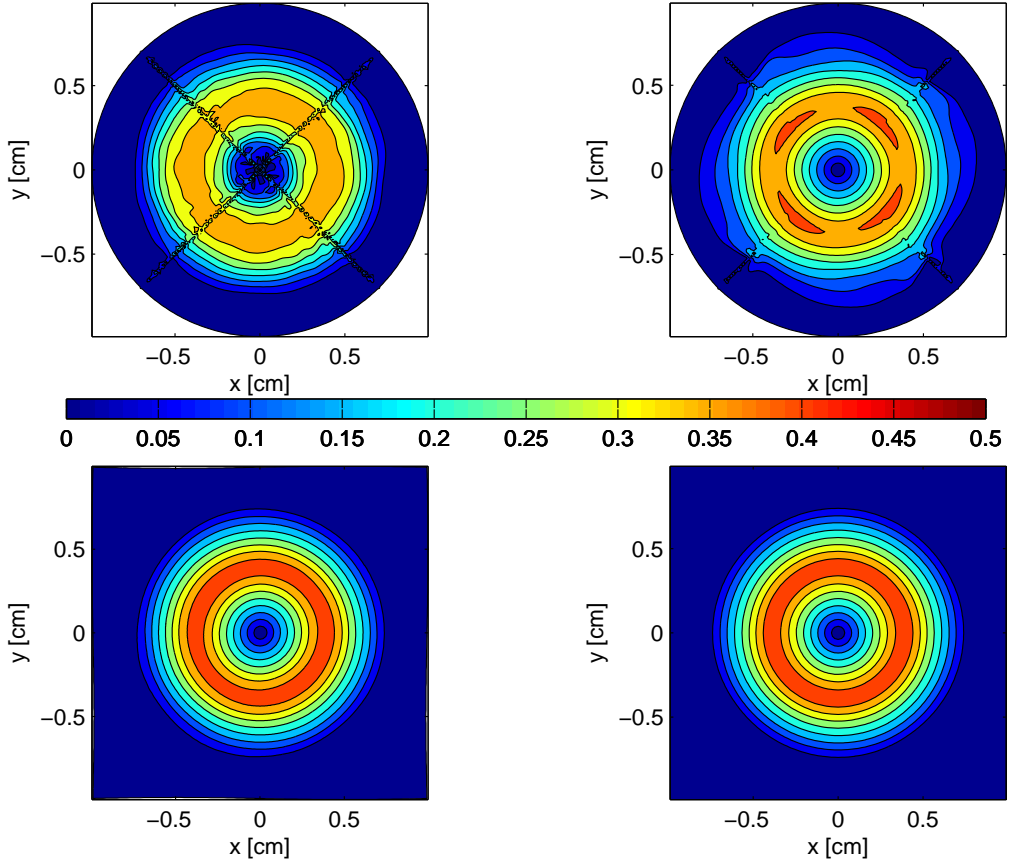}
\caption{Mach number in the Gresho vortex test after $2\,{\rm s}$ with the WENO finite difference 
         scheme with initial Mach number of $0.5$.}\label{fig-GreshoMa05}
\end{figure}

The Mach number ${\rm Ma}_{\rm ref}=0.1$ of this test is rather low for an explicit time 
integration scheme, but \citet{HappenhoferGrimm-StreleKupkaetal2013} showed that the WENO 
scheme with Runge--Kutta time integration yields reliable results in this regime. As shown in 
Figure~\ref{fig-GreshoMa05}, the deformations due to the grid get smaller with higher Mach 
numbers, but in any case, the results with the finite volume scheme are more accurate.
On the other hand, this explains why applying the mapped grid technique to the WENO finite 
difference scheme in \citet{Shu2003} worked properly: the mapping function were smooth, and
the Mach number in the numerical tests was always larger than $1$.

We conclude that the WENO finite difference scheme should only be used in combination with
the mapped grid technique if the mapping function is smooth and if the Mach number of the 
simulation is high. The reason for the bad performance lies in the violation of the preservation 
of the freestream. The finite volume scheme, however, yields accurate results even on 
non-smooth grids and in low Mach number tests.

\subsection{Sod Shock Tube}

With the Sod Shock Tube \citep{Sod1978} we test the performance of our schemes in the high 
Mach number regime. The computational domain is $\left[ -0,5, 0.5 \right]^2$ and $\gamma = 1.4$. 
In each direction, we use $100$ grid points. The inital shock position is at $0.2$ in 
$x$ direction.

\begin{figure}
\vspace{0pt}
\centering
\includegraphics[width=1.0\textwidth]{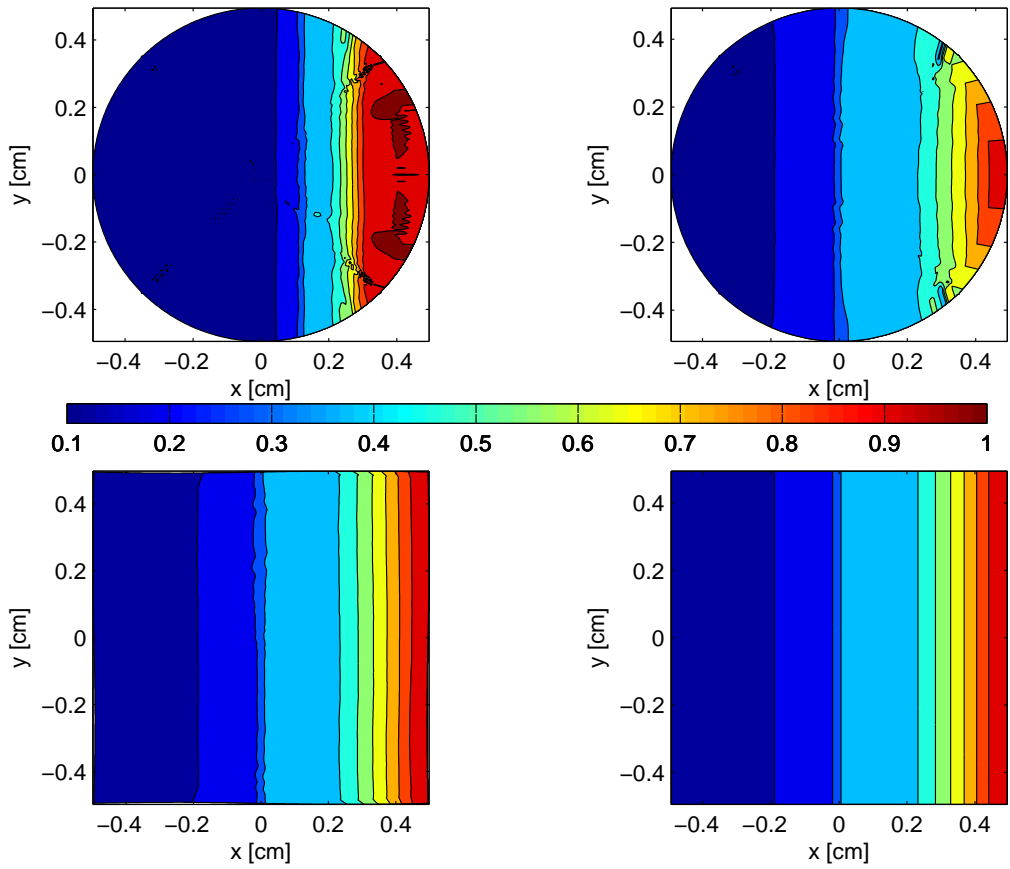}
\caption{Density in the Sod Shocktube Test after $0.22\,{\rm s}$ with the WENO finite difference scheme.
$M_1$ (top left panel) after $0.09\,{\rm s}$. In the outermost three layers, outflow boundary
conditions are set.}
\end{figure}

\begin{figure}
\vspace{0pt}
\centering
\includegraphics[width=1.0\textwidth]{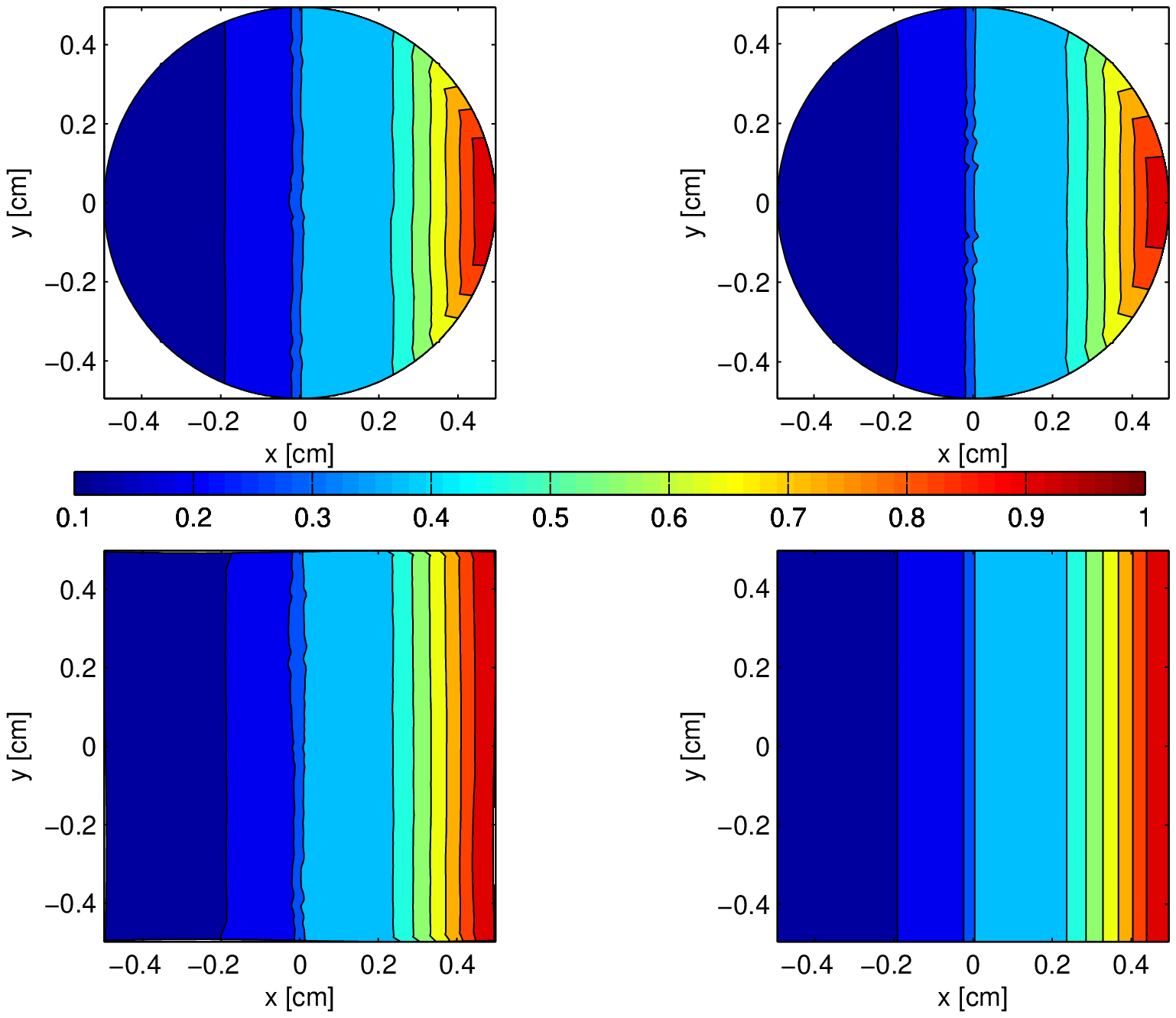}
\caption{Density in the Sod Shocktube Test after $0.22\,{\rm s}$ with the WENO finite volume scheme. 
In the outermost three layers, outflow boundary conditions are set.}
\end{figure}

Even in this setup where the Mach number is rather high, the WENO finite difference scheme 
performs badly with the non-smooth mapping functions $M_1$ and $M_2$. With $M_1$, the simulation
crashed after $0.09\,{\rm s}$ because of negative densities. At the diagonals, numerical artifacts 
occur even without any flow present at this position as a consequence of the violation of 
the freestream preservation.

The simulations with the finite volume scheme do not show any anomalies. On the smooth 
grids, the results from the two schemes are very similar.

\subsection{Nonlinear Advection}

\citet{YeeSandhamDjomehri1999}, \citet{ZhangZhangShu2011} and \citet{KifonidisMueller2012}
suggested a non-linear flow example to test the order of accuracy of a scheme. They 
showed that a smooth linear flow is not a suitable test case to test the accuracy of a 
finite volume method, because for a linear initial condition, the integration used to 
transform equation~\eqref{eq-curvdisc1} into~\eqref{eq-curvdisc2} resp.\ in 
equation~\eqref{eq-FVInt} is exact, and the overall order of the method is not 
restricted. They suggested to instead use the non-linear initial conditions 

\begin{subequations}
\begin{equation}
  \rho_{\rm mean} = 1,\,u_{\rm mean} = 1,\,v_{\rm mean} = 1,\,p_{\rm mean} = 1,
\end{equation}

\noindent with the perturbations of the velocities $u$ and $v$, the temperature $T \sim p/\rho$,
and the entropy $S \sim p/\rho^{1.4}$

\begin{equation}
  ( \delta u, \delta v ) = \frac{\epsilon}{2 \pi} \mathrm{e}^{0.5(1-r^2)}(-\overline{y},\overline{x}),\,
  \delta T = -\frac{(\gamma-1) \epsilon^2}{8 \gamma \pi^2} \mathrm{e}^{1-r^2},\,
  \delta S = 0.
\end{equation}
\end{subequations}

The computational domain is $\left[ 0, 10 \right]^2$, $(\overline{x},\overline{y}) = ( x-5, y-5 )$,
$r^2 = \overline{x}^2 + \overline{y}^2$, $\gamma=1.4$ and the vortex strength $\epsilon$ is $5$. 
The analytical solution is passive convection of the vortex with the mean flow. For the explicit 
time integration, we used the third-order TVD3 scheme \citep{ShuOsher1988} in this test.

\begin{figure}
\vspace{0pt}
\centering
\includegraphics[width=0.75\textwidth]{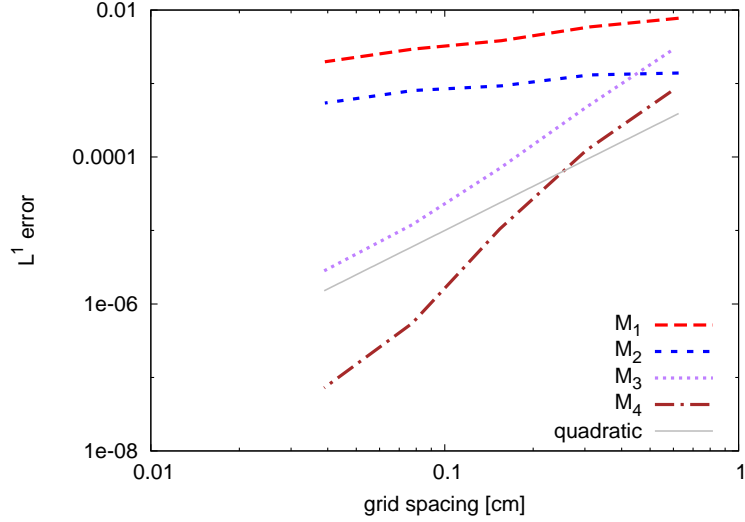}
\caption{Order of accuracy of the WENO finite difference scheme for the 
nonlinear advection problem measured in the $L^1$ norm. The grey line indicates
second-order convergence.}\label{fig-FDOoA}
\end{figure}

\begin{figure}
\vspace{0pt}
\centering
\includegraphics[width=0.75\textwidth]{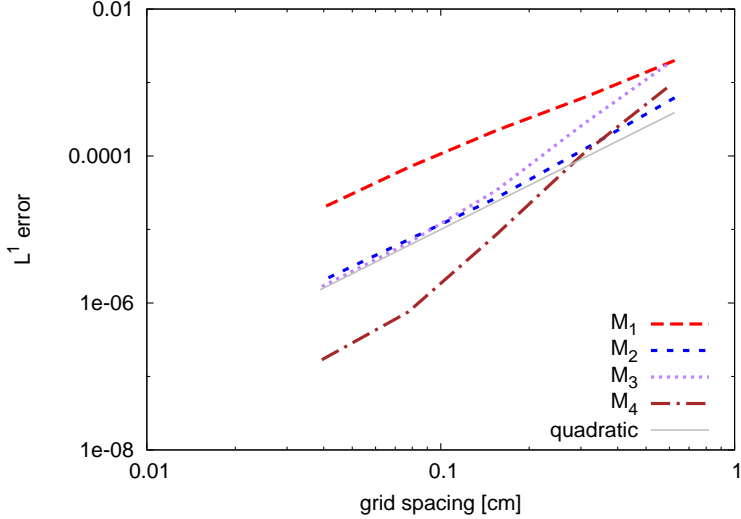}
\caption{Order of accuracy of the WENO finite volume scheme for the 
nonlinear advection problem measured in the $L^1$ norm. The grey line indicates
second-order convergence.}\label{fig-FVOoA}
\end{figure}

In Figures~\ref{fig-FDOoA} and \ref{fig-FVOoA}, the error of the WENO finite difference and
the finite volume scheme for the nonlinear advection problem are shown. The error is measured
by comparing the numerical solution after $0.2\,{\rm s}$ to the analytical one in the $L^1$ norm. 
We conclude that for the Cartesian and the smooth grid defined by the mapping functions $M_4$ and
$M_3$, both schemes show comparable error sizes. The empirical order of convergence is between two 
and three for $M_3$ and higher than three for $M_4$ in both cases.

For the non-smooth mapping functions $M_1$ and $M_2$, the error does not decrease significantly 
for the finite difference scheme, whereas first- to second-order convergence can be observed 
with the finite volume scheme.

\begin{table}[t]
\begin{center}
\begin{tabular}{c|ll|ll|ll|ll}
grid    & \multicolumn{2}{c}{$M_1$}  & \multicolumn{2}{|c}{$M_2$}  
        & \multicolumn{2}{|c}{$M_3$} & \multicolumn{2}{|c}{$M_4$}  \\
spacing & error   & order & error   & order & error   & order & error   & order  \\
\hline
6.25e-1 & 7.74e-3 &       & 1.39e-3 &       & 2.77e-3 &       & 7.74e-4 &       \\
3.13e-1 & 5.89e-3 & 0.392 & 1.31e-3 & 0.088 & 4.76e-4 & 2.539 & 1.23e-4 & 2.652 \\ 
1.56e-1 & 3.83e-3 & 0.621 & 9.29e-4 & 0.492 & 7.01e-5 & 2.764 & 1.05e-5 & 3.548 \\
7.81e-2 & 2.95e-3 & 0.380 & 8.03e-4 & 0.209 & 1.20e-5 & 2.546 & 5.46e-7 & 4.270 \\
3.91e-2 & 1.98e-3 & 0.572 & 5.43e-4 & 0.566 & 2.83e-6 & 2.087 & 7.21e-8 & 2.919
\end{tabular}
\caption{Mean $L_1$ error sizes and order of accuracy for the finite difference scheme.}
\label{tab-errFD}
\end{center}
\end{table}

\begin{table}[t]
\begin{center}
\begin{tabular}{c|ll|ll|ll|ll}
grid    & \multicolumn{2}{c}{$M_1$}  & \multicolumn{2}{|c}{$M_2$}  
        & \multicolumn{2}{|c}{$M_3$} & \multicolumn{2}{|c}{$M_4$}  \\
spacing & error   & order & error   & order & error   & order & error   & order \\
\hline
6.25e-1 & 2.00e-3 &       & 6.20e-4 &       & 1.74e-3 &       & 8.46e-4 &       \\
3.13e-1 & 6.40e-4 & 1.643 & 1.27e-4 & 2.283 & 2.64e-4 & 2.717 & 1.04e-4 & 3.018 \\ 
1.56e-1 & 2.26e-4 & 1.498 & 2.73e-5 & 2.222 & 3.37e-5 & 2.970 & 8.40e-6 & 3.635 \\
7.81e-2 & 7.06e-5 & 1.681 & 7.32e-6 & 1.900 & 6.53e-6 & 2.369 & 7.53e-7 & 3.480 \\
3.91e-2 & 1.93e-5 & 1.869 & 1.97e-6 & 1.894 & 1.63e-6 & 1.998 & 1.64e-7 & 2.201
\end{tabular}
\caption{Mean $L_1$ error sizes and order of accuracy for the finite volume scheme.}
\label{tab-errFV}
\end{center}
\end{table}

In the finite volume case, we expect second-order convergence for all mappings 
due to the second-order integration formula used to obtain equation~\eqref{eq-curvdisc2}.
Consequently, we observe second-order convergence for all grids at least asymptotically 
for high resolution in Table~\ref{tab-errFV}. For low resolutions, the numerical solution 
exhibits third-order convergence indicating that the temporal error dominates in this 
regime. We note that the Courant number is $0.1$ in all of our tests minimising the error 
due to the time integration scheme. We conclude that the absolute magnitude of the spatial 
error strongly depends on the mapping function, whereas the convergence order is 
restricted by the integration rule used to convert equation~\eqref{eq-curvdisc1} 
into~\eqref{eq-curvdisc2}. To obtain a higher than second order accurate scheme, 
high-order integration formulae must be used \citep{CasperAtkins1993,ZhangZhangShu2011}.

On Cartesian grids, the finite difference algorithm is superior compared to the 
finite volume algorithm. As expected, the scheme is at least third order accurate for 
all resolutions whereas the finite volume scheme shows second-order convergence 
for high resolutions. For the smooth mapping $M_3$, \citet{VisbalGaitonde2002} showed 
that the violation of the freestream preservation leads to large errors which dominate 
the overall error and decrease the convergence speed. With the non-smooth functions $M_1$ 
and $M_2$, this error is particularly large at the diagonals as we can observe for the 
Gresho vortex in Figure~\ref{fig-GreshoMa01-FD}. This error does hardly decrease with grid 
spacing and leads to the slow convergence found in the data from Table~\ref{tab-errFD}.
We can explain this slow convergence by the fact that in the finite difference approach,
the fluxes ${\bf \hat F}$ and ${\bf \hat G}$ are reconstructed. But these fluxes are 
non-smooth themselves since they contain the non-smooth metric terms. Therefore, the 
reconstruction process is only of low order, too.

This test demonstrates once more that the WENO finite difference scheme should only 
be combined with the mapped grid technique if the mapping function is smooth. With 
the finite volume scheme, the simulation converges also on non-smooth grids, but 
at a much slower rate than on smooth grids. With both algorithms, the ``smoother'' 
function $M_2$ yields more accurate results than $M_1$. We conclude that also the 
finite volume scheme performs better the smoother the grid is. Therefore, non-smooth 
grids should only be used if absolutely necessary.

For coarse resolutions, the results of $M_2$ with the finite volume scheme are better 
than the results with $M_1$, but also with the smooth mapping $M_3$. They are even 
comparable with the Cartesian mapping $M_4$. In astrophysical applications where 
the grid resolution usually is very coarse, grids like the one defined by $M_2$ 
combined with the WENO finite volume algorithm may well yield sufficiently accurate 
results. These grids are an acceptable alternative, if a non-smooth grid is needed 
by the problem geometry.

\section{Discussion and Outlook}
\label{sec-discussion}

Curvilinear coordinates are an easy and efficient way to extend existing codes 
written for Cartesian grids to more complicated geometries. The grid can be generated 
either by a (analytical or numerical) mapping function, or by an external grid 
generation program. Within the framework introduced in Section~\ref{sec-mapped}
it must always be structured.

For the case of spherical domains, \citet{Calhounetal2008} provided mapping functions
which are not strongly differentiable. We showed that the analytical transformation 
also works in a weak sense without assuming strong differentiability of the mapping 
function.

In \citet{Shu2003}, the application of the WENO finite difference scheme to smooth
curvilinear grids is shown. The results are of high accuracy, since the mapping 
functions are smooth and the Mach number of the numerical tests are high.

In this paper, for the first time the WENO finite difference and finite volume scheme 
were applied with non-smooth mapping functions as those defined by the mapping 
functions from \citet{Calhounetal2008}. Particular attention is paid to problems 
arising when the Mach number of the flow is rather low. Whereas the WENO finite volume 
scheme performs well in these cases, the WENO finite difference scheme does not give a 
convergent solution. It only works if the mapping function is smooth, but even in this 
case the finite volume scheme yields better results for the low Mach number tests 
presented in this paper.

Since it is only a minor algorithmic change from the WENO finite difference to the 
WENO finite volume scheme, we recommend to switch to the WENO finite volume scheme 
when the mapped grid technique is used for non-smooth mapping functions and also for 
low Mach numbers. As demonstrated in Section~\ref{sec-results} the smoother the mapping 
function is, the more accurate are the results. Therefore, one should always use the 
smoothest mapping function allowed by the simulation setup.

In theory, the finite volume scheme is only second order accurate. In our calculations,
however, the empirical order of accuracy of the finite volume scheme was higher. To
increase the theoretical order of accuracy the computational requirements and 
the complexity of the code has to be increased considerably \citep{ZhangZhangShu2011}.
In practice, WENO schemes usually are combined with lower order Runge--Kutta methods
for time integration with highest-possible Courant numbers. The overall error of the 
scheme will be dominated by the temporal error, and the overall order of the method 
will be limited to two or three, anyway.

Furthermore, in astrophysical simulations the typical resolution is rather coarse. In 
this case the spatial error dominates and fast second order time integrators such as 
SSP\,RK(3,2) (\citealt{Kraaijevanger1991}, cf.\ also 
\citealt{KetchesonMacdonaldGottlieb2009,KupkaHappenhoferHiguerasKoch2012}) are 
sufficient. In this regime, non-smooth mapping functions perform nearly as well as 
smooth ones provided they are combined with the WENO finite volume scheme. We conclude 
that they are an acceptable alternative, if the problem geometry requires the use of 
non-smooth mapping functions. Moreover curvilinear coordinates provide enough 
flexibility for most problems in computational astrophysics whereas keeping the high
efficiency and accuracy of the Cartesian schemes they are based on.

We show how the grid capability of an existing code written for Cartesian coordinates
can be extended to more general geometries. In this way, a wide variety of astrophysical 
problems can be tackled with ANTARES which were not feasible with standard grids in 
numerical astrophysics and without any fundamental changes in the numerical method.

In the near future, we plan to extend our work to the Navier--Stokes equations with 
gravity and diffusion in three spatial dimensions. It would be interesting to apply 
low Mach number methods such as the one presented in \citet{KwatraSuGretarssonFedkiw2009}
and \citet{HappenhoferGrimm-StreleKupkaetal2013} to curvilinear coordinates and further
improve their performance for low Mach numbers.

Furthermore, the influence of the Mach number on the results needs additional 
investigations. We assume that the distortions due to the freestream preservation 
violation for finite difference schemes are hidden when there are fast motions 
of the fluid. Only with low Mach numbers, the numerical errors get large enough 
to disturb the numerical solution considerably.

\section*{Acknowledgements}

We acknowledge financial support from the Austrian Science fund (FWF), projects 
P21742 and P25229. HGS wants to thank E.~M\"uller, A.~Wongwathanarat, P.~Edelmann 
and F.~Miczek for helpful and inspiring discussions and the MPA Garching for 
a grant for two research stays in Garching. Particular gratitude is owed to 
A.~Wongwathanarat for adverting the work of Calhoun et al.\ to HGS.

\appendix

\section{Finite difference and finite volume discretisation of the Euler equations}
\label{app-disc}

First, we derive the finite difference and finite volume discretisation of the Euler 
equations for the one-dimensional case. The differential form of the Euler equations 
in one spatial dimension is

\begin{equation}
  \frac{\partial}{\partial t}{\bf Q} 
        + \frac{\partial}{\partial x}{\bf F} = 0,\label{eulerdiff1d}
\end{equation}

\begin{equation}
  {\bf Q} = \left( \begin{array}{c}
                      \rho   \\
                      \rho u \\
                      E
                   \end{array} \right),\ 
  {\bf F} \left( {\bf Q} \right) = \left( \begin{array}{c}
                                                 \rho u       \\
                                                 \rho u^2 + p \\
                                                 ( p + E ) u
                                          \end{array} \right),
\end{equation}

\noindent where $p=p(\rho,e)$. The internal energy $e$ is given by
$e = E - \frac{u^2}{2 \rho}$. Let $\Omega = \left[a, b\right]$ 
and let a grid on $\Omega$ be defined on the half-integer nodes, i.e.

\begin{equation}
  x_{i+\frac{1}{2}} = x_{i-\frac{1}{2}} + \delta x_i,\ i=1,\hdots,n,\ x_{\frac{1}{2}}=a, x_{n+\frac{1}{2}}=b.\label{grid}
\end{equation}

The grid is not necessarily equidistant. The index $i$ refers to the centre of the 
cell $C_i=\left[ x_{i-\frac{1}{2}}, x_{i+\frac{1}{2}} \right]$.

\begin{definition} 
The {\bf Cell Average Operator} $\mathrm{A}$ on a grid \eqref{grid} is defined by

\begin{equation}
  \mathrm{A}(h)_i:=\frac{1}{\delta x_i} \int_{x_{i-\frac{1}{2}}}^{x_{i+\frac{1}{2}}} h(\hat{x})d\hat{x}.
\end{equation}

\end{definition}

Applying $\mathrm{A}$ to \eqref{eulerdiff1d} gives

\begin{equation}
  \frac{\partial}{\partial t}\left( \mathrm{A}{\bf Q} \right)_i + \frac{\delta {\bf F}_i}{\delta x_i} = 0 
     \text{ with } \frac{\delta {\bf F}_i}{\delta x_i} 
          = \frac{{\bf F}_{i+\frac{1}{2}}-{\bf F}_{i-\frac{1}{2}}}{\delta x_i}\label{startFV}
\end{equation}

\noindent with ${\bf F}_{i+\frac{1}{2}}={\bf F} \left( {\bf Q}(x_{i+\frac{1}{2}}) \right)$. 
$\left( \mathrm{A}{\bf Q} \right)_i=:\overline{{\bf Q}}_i$ is the {\it Cell Average} of ${\bf Q}$ 
in the cell $C_i=\left[x_{i-\frac{1}{2}},x_{i+\frac{1}{2}}\right]$.

Applying now $\mathrm{A}^{-1}$ as in \citet{Merriman2003}, we get

\begin{equation}
  \frac{\partial}{\partial t}{\bf Q}_i+ \mathrm{A}^{-1}\frac{\delta {\bf F}_i}{\delta x_i} = 0.
  \label{eq-befstrFD}
\end{equation}

\citet{Merriman2003} showed that

\begin{equation}
  \mathrm{A}^{-1}\frac{\delta {\bf F}_i}{\delta x_i} 
  = \frac{\delta \left( \mathrm{A}^{-1}{\bf F}_i \right) }{\delta x_i}
  \text{ if the grid is equidistant, i.e. }\delta x_i=\delta x\ \forall i.
\end{equation}

In general, this is not the case since $\frac{\delta {\bf F}_i}{\delta x_i}$ is a 
translation operator, but $\mathrm{A}$ is not translation-invariant. For an equidistant
grid, however, \eqref{eq-befstrFD} is equivalent to

\begin{equation}
  \frac{\partial}{\partial t}{\bf Q}_i + \frac{\delta {\bf f}_i}{\delta x_i} 
  = 0 \text{  with }{\bf F}_i=(\mathrm{A}{\bf f})_i.\label{startFD}
\end{equation}

\eqref{startFD} is called the {\bf Shu--Osher form} of~\eqref{eulerdiff1d}. It is equivalent 
to~\eqref{eulerdiff2d} if the grid is equidistant \citep{Merriman2003}. ${\bf f}$ is defined
such that its average in cell $C_i$ is given by the point value of the analytical flux function
${\bf F}$ at the cell centre. In terms of \citet{ShuOsher1988}, ${\bf F}$ is exactly the
``primitive function'' of ${\bf f}$.

A finite volume scheme starts with equation~\eqref{startFV} and computes approximations for 
${\bf Q}_{i+\frac{1}{2}}$ from the given cell averaged $\overline{{\bf Q}}_i$. The numerical 
flux ${\bf F}_{i+\frac{1}{2}}$ is calculated by

\begin{equation}
  {\bf F}_{i+\frac{1}{2}} = {\bf F} \left( {\bf Q}_{i+\frac{1}{2}} \right).
\end{equation}

In contrast, the finite difference scheme starts with equation~\eqref{startFD} and computes 
approximations for ${\bf f}_{i+\frac{1}{2}}$ from the values of the analytical flux function 
${\bf F}$. Since ${\bf F}_i=(\mathrm{A}{\bf f})_i$, both approaches are computationally
equivalent and require the numerical solution of the following

\begin{problem}[Reconstruction problem]\label{reconstructprob}
  Given a set of cell averages, compute the value of the underlying 
  function at the half-integer nodes.
\end{problem}

\begin{definition} 
The {\bf Reconstruction Operator} $\mathrm{R}$ acts on cell averages and 
reconstructs the point value of the underlying function

\begin{equation}
  \mathrm{R}(\mathrm{A}h)_{i+\frac{1}{2}}=h_{i+\frac{1}{2}}
\end{equation}

\noindent on a given grid~\eqref{grid}.
\end{definition}

If the grid is equidistant, the same algorithm can be used to compute ${\bf Q}_{i+\frac{1}{2}}$ 
and ${\bf f}_{i+\frac{1}{2}}$. The only difference lies in the input for the reconstruction 
process: in the case of a finite volume scheme, the inputs are given by the cell averages 
$\overline{\bf Q}_i$ of the state vector, whereas in the case of a finite difference scheme, 
they are given by the analytical flux function ${\bf F}$ evaluated at the cell centre.

Given a reconstruction operator $\mathrm{R}$, Problem~\ref{reconstructprob} can be solved. 
The WENO reconstruction operator is described in detail in \citet{Shu2003} and also 
in~\ref{app-rec5th}, following the cited reference. In general, finite volume methods can
be designed on non-equidistant grids. However, since the WENO reconstruction operator as
it is described in~\ref{app-rec5th} works only on equidistant grids, the applicability of
the finite volume method using this reconstruction method is restricted to equidistant grids,
too.

The advantage of the finite difference scheme lies in its easy extension to higher 
dimensions. Let a region (i.e., a non-empty, open, and connected set) $\Omega \subset \R^2$ 
be given. The Euler equations are a system of partial differential equations 
on $\Omega$. In two spatial dimensions and in a Cartesian coordinate system, 
their differential form is

\begin{equation}
  \frac{\partial}{\partial t}{\bf Q} 
        + \frac{\partial}{\partial x}{\bf F} 
        + \frac{\partial}{\partial y}{\bf G} = 0, \label{app-eulerdiff2d}
\end{equation}

\noindent with the state vector ${\bf Q}$ and the flux functions ${\bf F}$ and 
${\bf G}$ given by

\begin{equation}
  {\bf Q} = \left( \begin{array}{c}
                      \rho   \\
                      \rho u \\
                      \rho v \\
                      E
                   \end{array} \right),
  {\bf F} \left( {\bf Q} \right) = \left( \begin{array}{c}
                                                 \rho u       \\
                                                 \rho u^2 + p \\
                                                 \rho u v     \\
                                                 ( p + E ) u
                                          \end{array} \right),
  {\bf G} \left( {\bf Q} \right) = \left( \begin{array}{c}
                                                 \rho v       \\
                                                 \rho v u     \\
                                                 \rho v^2 + p \\
                                                 ( p + E ) v
                                          \end{array} \right),
\end{equation}

\noindent where $p=p(\rho,e)$. In the following, we will write ${\bf u} := \left( u, v \right)^T$
for the velocity vector.

In the finite volume approach, applying $\mathrm{A}_x \mathrm{A}_y$ to the two-dimensional 
Euler equations~\eqref{eulerdiff2d} results in

\begin{equation}
    \frac{\partial}{\partial t} \left( \mathrm{A}_x \mathrm{A}_y {\bf Q} \right)_{i,j} 
  + \frac{\mathrm{A}_y({\bf F}_{i+\frac{1}{2}})_j-\mathrm{A}_y({\bf F}_{i-\frac{1}{2}})_j}{\delta x} 
  + \frac{\mathrm{A}_x({\bf G}_{j+\frac{1}{2}})_i-\mathrm{A}_x({\bf G}_{j-\frac{1}{2}})_i}{\delta y} = 0.\label{FV2d}
\end{equation}

The fluxes are now line integrals over the cell boundaries. With the midpoint rule,

\begin{equation}
  \mathrm{A}_y({\bf F}_{i+\frac{1}{2}})_j = {\bf F}_{i+\frac{1}{2},j} + O(h^2).\label{eq-FVInt}
\end{equation}

With only one evaluation of the fluxes, the overall order of the method is restricted to two.
We remark that if ${\bf F}$ is linear, this integration is exact.

Contrary, in the finite difference approach, \eqref{app-eulerdiff2d} is transformed to

\begin{equation}
    \frac{\partial}{\partial t}{\bf Q}_{i,j} 
  + \frac{{\bf f}_{i+\frac{1}{2},j}-{\bf f}_{i-\frac{1}{2},j}}{\delta x} 
  + \frac{{\bf g}_{i,j+\frac{1}{2}}-{\bf g}_{i,j-\frac{1}{2}}}{\delta y}= 0
\end{equation}

\noindent with ${\bf F}=\mathrm{A}_x({\bf f})$ and ${\bf G}=\mathrm{A}_y({\bf g})$. The 
overall order of the method only depends on the order of the reconstruction of ${\bf f}$ 
and ${\bf g}$, which are one-dimensional reconstruction problems. The one-dimensional 
algorithms can be directly applied without loss of order of accuracy.

Even if the high order one-dimensional WENO reconstruction algorithm is applied to 
evaluate the flux functions in~\eqref{FV2d}, the resulting finite volume method is only
second order accurate. To obtain a truely high-order multidimensional finite volume 
method, more complicated integrations for the line integrals in~\eqref{FV2d} must be 
performed, increasing the computational costs of the finite volume scheme tremendously 
\citep{Shu2003,ColellaDorrHittingerMartin2011,ZhangZhangShu2011}. Therefore, finite 
difference schemes are clearly preferable on Cartesian grids.

\section{Classical Transformation to Curvilinear Coordinates Assuming Strong Differentiability}
\label{app-strong}

If there is a differentiable function 

\begin{equation}
  M:\ \left[ -1, 1 \right]^2 \to \Omega,\ M(\xi,\eta) = (x, y)^T,
\end{equation}

\noindent we can transform the Euler equations in differential form~\eqref{eulerdiff2d} to

\begin{subequations}
\begin{equation}
  \frac{\partial}{\partial t}J^{-1}{\bf Q} + \frac{\partial}{\partial \xi }{\bf \hat{F}} 
                                           + \frac{\partial}{\partial \eta}{\bf \hat{G}} = 0 
\end{equation}

\noindent with

\begin{align}
  {\bf \hat{F}} = &   \frac{\partial y}{\partial \eta} {\bf F} - \frac{\partial x}{\partial \eta} {\bf G}, \\
  {\bf \hat{G}} = & - \frac{\partial y}{\partial \xi}  {\bf F} + \frac{\partial x}{\partial \xi}  {\bf G},
\end{align}

\noindent and the determinant of the inverse Jacobian of the transformation

\begin{equation}
  J^{-1} = \left| \frac{\partial (x,y)}{\partial (\xi,\eta)} \right|
         = \frac{\partial y}{\partial \eta} \frac{\partial x}{\partial \xi} 
         - \frac{\partial y}{\partial \xi} \frac{\partial x}{\partial \eta}.
\end{equation} \label{eq-curvi2dB}
\end{subequations}

In this way, the conservation law~\eqref{eulerdiff2d} defined in the physical space is transformed
into a conservation law in the computational space. If $M$ is at least in $C^1$, all derivatives 
and the Jacobian are well-defined \citep{Vinokur1974,KifonidisMueller2012}.

Following the description in \citet{TannehillAndersonPletcher1997}, this form can be derived by
multiplying \eqref{eulerdiff2d} with $J^{-1}$ and rearranging terms. First we look at 
$\frac{\partial {\bf F}}{\partial x} J^{-1}$. With the chain rule of differentiation,

\begin{align*}
   \pad[{\bf F}]{x} J^{-1} & = \left( \pad[\xi ]{x} \pad[{\bf F}]{\xi } 
                             + \pad[\eta]{x} \pad[{\bf F}]{\eta} \right) J^{-1} \\
                           & = \pad{\xi } \left( {\bf F} \pad[\xi ]{x} J^{-1} \right)
                             + \pad{\eta} \left( {\bf F} \pad[\eta]{x} J^{-1} \right) \\
                           & - {\bf F} \pad{\xi } \left( \pad[\xi ]{x} J^{-1} \right)
                             - {\bf F} \pad{\eta} \left( \pad[\eta]{x} J^{-1} \right).
\end{align*}

In two dimensions,

\begin{equation}
    \left( \begin{array}{cc}
             \pad[\xi ]{x} & \pad[\xi ]{y} \\
             \pad[\eta]{x} & \pad[\eta]{y}
           \end{array} \right)
  = \left( \begin{array}{cc}
             \pad[x]{\xi} & \pad[x]{\eta} \\
             \pad[y]{\xi} & \pad[y]{\eta}
           \end{array} \right)^{-1}
  = J \left( \begin{array}{cc}
                 \pad[y]{\eta} & - \pad[x]{\eta} \\
               - \pad[y]{\xi } &   \pad[x]{\xi }
             \end{array} \right),
\end{equation}

\noindent and as a direct consequence,

\begin{equation}
  \pad[\xi ]{x} J^{-1} =   \pad[y]{\eta},\ \pad[\xi ]{y} J^{-1} = - \pad[x]{\eta},\ 
  \pad[\eta]{x} J^{-1} = - \pad[y]{\xi },\ \pad[\eta]{y} J^{-1} =   \pad[x]{\xi }.
\end{equation}

We can further write

\begin{align*}
   \pad[{\bf F}]{x} J^{-1} & = \pad{\xi } \left(   \pad[y]{\eta} {\bf F} \right)
                             + \pad{\eta} \left( - \pad[y]{\xi } {\bf F} \right)
                             - {\bf F} \underbrace{\left( \frac{\partial^2 y}{\partial \xi \partial \eta} 
                                                        - \frac{\partial^2 y}{\partial \eta \partial \xi} \right)}_{=0}.
\end{align*}

Here we assumed the mapping function to be at least in $C^2$. The same procedure leads 
to similar expressions for $\frac{\partial {\bf G}}{\partial y} J^{-1}$.
Plugging all these expressions into \eqref{eulerdiff2d} leads to \eqref{eq-curvi2dB}.

\section{The $5$th order WENO Reconstruction Algorithm}
\label{app-rec5th}

In the following, the WENO reconstruction operator is derived following 
\citet{Shu2003}. The purpose of the reconstruction operator is to solve 
Problem~\ref{reconstructprob}, i.e.\ reconstruct the value of the underlying 
function at a certain position given a set of cell averages.

We will only consider equidistant one-dimensional grids and reconstruction 
of the value at the half-integer node $i+\frac{1}{2}$. This is sufficient 
for the finite difference and finite volume algorithms used in this paper.

The idea of the WENO reconstruction process is to use several stencils in 
the neighbourhood of $i+\frac{1}{2}$. On each of the stencils, an interpolating 
polynomial of high order is defined. The interpolated value is obtained by summing 
these polynomials weighting them according to their smoothness. If a discontinuity 
is contained in the stencil of a polynomial, its weight will be very small thereby 
avoiding oscillatory behaviour known from high order interpolation. 

Assume that the cell averages $\overline{\phi}_i=(\mathrm{A}\phi)_i$ of the function $\phi$ 
are given and $\phi_{i+\frac{1}{2}}$ should be reconstructed. We consider $k$ stencils

\begin{equation}
  S_r(i) = \{x_{i-r},\hdots,x_{i-r+k-1} \},\ r=0,\hdots,k-1.
\end{equation}

On each stencil $S_r(i)$ a polynomial $p_r$ of degree $k-1$ is defined such that
the approximation $\phi_{i+\frac{1}{2}}^{(r)}$ to $\phi_{i+\frac{1}{2}}$ is given by

\begin{equation}
  \phi_{i+\frac{1}{2}}^{(r)} = p_r(x_{i+\frac{1}{2}}) + O((\delta x)^k)
\end{equation}

\noindent and

\begin{equation}
  \mathrm{A}(p_r) = \mathrm{A}(\phi) = \overline{\phi} \text{ on }S_r(i).
\end{equation}

Solving the resulting linear system for the case $k=3$ gives the interpolation polynomials

\begin{align}
  p_0(x_{i+\frac{1}{2}}) & =   \frac{1}{3}  \overline{\phi}_{i-2} 
                             - \frac{7}{6}  \overline{\phi}_{i-1} 
                             + \frac{11}{6} \overline{\phi}_{i},   \\
  p_1(x_{i+\frac{1}{2}}) & = - \frac{1}{6}  \overline{\phi}_{i-1} 
                             + \frac{5}{6}  \overline{\phi}_{i}   
                             + \frac{1}{3}  \overline{\phi}_{i+1}, \\
  p_2(x_{i+\frac{1}{2}}) & =   \frac{1}{3}  \overline{\phi}_{i}   
                             + \frac{5}{6}  \overline{\phi}_{i+1} 
                             - \frac{1}{6}  \overline{\phi}_{i+2},
\end{align}

The Lagrange polynomial of fifth order can be obtained by combination of the three 
polynomials of third order. If we define the weights

\begin{equation}
  d_0 = \frac{3}{10}, d_1 = \frac{3}{5}, d_2 = \frac{1}{10},
\end{equation}

\noindent the fifth order interpolation polynomial $p^{(5)}$ can be obtained by

\begin{equation}
  p^{(5)}(x) = d_0 p_0(x) + d_1 p_1(x) + d_2 p_2(x).
\end{equation}

High-order polynomial interpolation is known to produce oscillatory results.
To avoid oscillations in the WENO approach, a convex combination of all candidate 
stencils $p_r$ is used to compute $\phi_{i+\frac{1}{2}}$. This procedure leads to 
non-oscillatory approximations of order $2k-1$, where $k$ is the width of each 
of the stencils $S_r(i)$.

Therefore, the approximation to $\phi_{i+\frac{1}{2}}$ is calculated by

\begin{equation}
  \phi_{i+\frac{1}{2}} = \omega_0 p_0(x_{i+\frac{1}{2}}) 
                       + \omega_1 p_1(x_{i+\frac{1}{2}}) 
                       + \omega_2 p_2(x_{i+\frac{1}{2}}),
\end{equation}

where $\omega_0$, $\omega_1$ and $\omega_2$ are nonlinear weights comparing the 
smoothness of the interpolation polynomials. Defining 

\begin{equation}
\begin{split}
  \beta_0 & = \frac{13}{12} \left( \overline{\phi}_{i}   -   \overline{\phi}_{i+1} 
                                 + \overline{\phi}_{i+2} \right)^2 
            + \frac{1}{4} \left( 3 \overline{\phi}_{i}   - 4 \overline{\phi}_{i+1} 
                                 +\overline{\phi}_{i+2} \right)^2, \\
  \beta_1 & = \frac{13}{12} \left( \overline{\phi}_{i-1} - 2 \overline{\phi}_{i}   
                                 + \overline{\phi}_{i+1} \right)^2 
            + \frac{1}{4} \left(   \overline{\phi}_{i-1} -   \overline{\phi}_{i+1} \right)^2, \\
  \beta_2 & = \frac{13}{12} \left( \overline{\phi}_{i-2} - 2 \overline{\phi}_{i-1} 
                                 + \overline{\phi}_{i}   \right)^2 
            + \frac{1}{4} \left(   \overline{\phi}_{i-2} - 4 \overline{\phi}_{i-1} \right)^2,
\end{split}
\end{equation}

\noindent we calculate

\begin{equation}
  {\tilde \omega}_0 = \frac{d_0}{(\beta_0 + \epsilon)^2},\ 
  {\tilde \omega}_1 = \frac{d_1}{(\beta_1 + \epsilon)^2},\ 
  {\tilde \omega}_2 = \frac{d_2}{(\beta_2 + \epsilon)^2},
\end{equation}

\noindent and finally

\begin{equation}
  \omega_0 = \frac{{\tilde \omega}_0}{{\tilde \omega}_0 + {\tilde \omega}_1 + {\tilde \omega}_2},\ 
  \omega_1 = \frac{{\tilde \omega}_1}{{\tilde \omega}_0 + {\tilde \omega}_1 + {\tilde \omega}_2},\ 
  \omega_2 = \frac{{\tilde \omega}_2}{{\tilde \omega}_0 + {\tilde \omega}_1 + {\tilde \omega}_2}.
\end{equation}

$\epsilon$ is a small parameter which is used to avoid division by zero.


\begin{thebibliography}{39}
\expandafter\ifx\csname natexlab\endcsname\relax\def\natexlab#1{#1}\fi
\providecommand{\bibinfo}[2]{#2}
\ifx\xfnm\relax \def\xfnm[#1]{\unskip,\space#1}\fi
\bibitem[{{C}alhoun et~al.(2008){C}alhoun, {H}elzel, and
  {L}e{V}eque}]{Calhounetal2008}
\bibinfo{author}{D.~A. {C}alhoun}, \bibinfo{author}{C.~{H}elzel},
  \bibinfo{author}{R.~J. {L}e{V}eque},
\newblock \bibinfo{title}{{L}ogically {R}ectangular {G}rids and {F}inite
  {V}olume {M}ethods for {PDE}s in {C}ircular and {S}pherical {D}omains},
\newblock \bibinfo{journal}{SIAM Review} \bibinfo{volume}{50}
  (\bibinfo{year}{2008}) \bibinfo{pages}{723 -- 752}.
\bibitem[{{B}rowning et~al.(2004){B}rowning, {B}run, and
  {T}oomre}]{BrowningBrunToomre2004}
\bibinfo{author}{M.~K. {B}rowning}, \bibinfo{author}{A.~S. {B}run},
  \bibinfo{author}{J.~{T}oomre},
\newblock \bibinfo{title}{{S}imulations {O}f {C}ore {C}onvection {I}n
  {R}otating {A}-type {S}tars: {D}ifferential {R}otation {A}nd {O}vershooting},
\newblock \bibinfo{journal}{ApJ} \bibinfo{volume}{601} (\bibinfo{year}{2004})
  \bibinfo{pages}{512 -- 529}.
\bibitem[{{C}ai et~al.(2011){C}ai, {C}han, and {D}eng}]{CaiChanDeng2011}
\bibinfo{author}{T.~{C}ai}, \bibinfo{author}{K.~L. {C}han},
  \bibinfo{author}{L.~{D}eng},
\newblock \bibinfo{title}{{N}umerical simulation of core convection by a
  multi-layer semi-implicit spherical spectral method},
\newblock \bibinfo{journal}{JCP} \bibinfo{volume}{230} (\bibinfo{year}{2011})
  \bibinfo{pages}{8698 -- 8712}.
\bibitem[{{E}vonuk and {G}latzmaier(2006)}]{EvonukGlatzmaier2006}
\bibinfo{author}{M.~{E}vonuk}, \bibinfo{author}{G.~A. {G}latzmaier},
\newblock \bibinfo{title}{2{D} study of the effects of the size of a solid core
  on the equatorial flow in giant planets},
\newblock \bibinfo{journal}{Icarus} \bibinfo{volume}{181}
  (\bibinfo{year}{2006}) \bibinfo{pages}{458 -- 464}.
\bibitem[{{E}vonuk and {G}latzmaier(2007)}]{EvonukGlatzmaier2007}
\bibinfo{author}{M.~{E}vonuk}, \bibinfo{author}{G.~A. {G}latzmaier},
\newblock \bibinfo{title}{{T}he effects of rotation rate on deep convection in
  giant planets with small solid cores},
\newblock \bibinfo{journal}{Planetary and Space Science} \bibinfo{volume}{55}
  (\bibinfo{year}{2007}) \bibinfo{pages}{407 -- 412}.
\bibitem[{{F}reytag et~al.(2002){F}reytag, {S}teffen, and
  {D}orch}]{FreytagSteffenDorch2002}
\bibinfo{author}{B.~{F}reytag}, \bibinfo{author}{M.~{S}teffen},
  \bibinfo{author}{B.~{D}orch},
\newblock \bibinfo{title}{{S}pots on the surface of {B}etelgeuse --- {R}esults
  from new 3{D} stellar convection models},
\newblock \bibinfo{journal}{Astron. Nachr.} \bibinfo{volume}{323}
  (\bibinfo{year}{2002}) \bibinfo{pages}{213 -- 219}.
\bibitem[{{Z}ingale et~al.(2013){Z}ingale, {N}onaka, {A}lmgren, {B}ell,
  {M}alone, and {O}rvedahl}]{ZingaleNonakaAlmgrenetal2013}
\bibinfo{author}{M.~{Z}ingale}, \bibinfo{author}{A.~{N}onaka},
  \bibinfo{author}{A.~S. {A}lmgren}, \bibinfo{author}{J.~B. {B}ell},
  \bibinfo{author}{C.~M. {M}alone}, \bibinfo{author}{R.~{O}rvedahl},
\newblock \bibinfo{title}{{L}ow {M}ach {N}umber {M}odeling of {C}onvection in
  {H}elium {S}hells on {S}ub-{C}handrasekhar {W}hite {D}warfs. {I}.
  {M}ethodology},
\newblock \bibinfo{journal}{ApJ} \bibinfo{volume}{764} (\bibinfo{year}{2013})
  \bibinfo{pages}{97 -- 110}.
\bibitem[{{M}ocz et~al.(2013){M}ocz, {V}ogelsberger, {S}ijacki, {P}akmor, and
  {H}ernquist}]{MoczVogelsbergerSijackietal2013}
\bibinfo{author}{P.~{M}ocz}, \bibinfo{author}{M.~{V}ogelsberger},
  \bibinfo{author}{D.~{S}ijacki}, \bibinfo{author}{R.~{P}akmor},
  \bibinfo{author}{L.~{H}ernquist},
\newblock \bibinfo{title}{{A} discontinuous {G}alerkin method for solving the
  fluid and {MHD} equations in astrophysical simulations},
\newblock \bibinfo{journal}{MNRAS}  (\bibinfo{year}{2013}).
  \bibinfo{note}{Available at \url{http://arxiv.org/abs/1305.5536}}.
\bibitem[{{M}uthsam et~al.(2010){M}uthsam, {K}upka, {L\"ow-Baselli},
  {O}bertscheider, {L}anger, and {L}enz}]{MuthsamKupkaLoew-Basellietal2010}
\bibinfo{author}{H.~J. {M}uthsam}, \bibinfo{author}{F.~{K}upka},
  \bibinfo{author}{B.~{L\"ow-Baselli}}, \bibinfo{author}{C.~{O}bertscheider},
  \bibinfo{author}{M.~{L}anger}, \bibinfo{author}{P.~{L}enz},
\newblock \bibinfo{title}{{ANTARES} -- {A} {N}umerical {T}ool for
  {A}strophysical {RES}earch with applications to solar granulation},
\newblock \bibinfo{journal}{NewA} \bibinfo{volume}{15} (\bibinfo{year}{2010})
  \bibinfo{pages}{460 -- 475}.
\bibitem[{{H}appenhofer et~al.(2013){H}appenhofer, {Grimm-Strele}, {K}upka,
  {L\"ow-Baselli}, and {M}uthsam}]{HappenhoferGrimm-StreleKupkaetal2013}
\bibinfo{author}{N.~{H}appenhofer}, \bibinfo{author}{H.~{Grimm-Strele}},
  \bibinfo{author}{F.~{K}upka}, \bibinfo{author}{B.~{L\"ow-Baselli}},
  \bibinfo{author}{H.~J. {M}uthsam},
\newblock \bibinfo{title}{{A} low {M}ach number solver: {E}nhancing
  applicability},
\newblock \bibinfo{journal}{JCP} \bibinfo{volume}{236} (\bibinfo{year}{2013})
  \bibinfo{pages}{96 -- 118}.
\bibitem[{{M}uthsam et~al.(2010){M}uthsam, {K}upka, {M}undprecht, {Z}aussinger,
  {Grimm-Strele}, and {H}appenhofer}]{MuthsamKupkaMundprechtetal2010}
\bibinfo{author}{H.~J. {M}uthsam}, \bibinfo{author}{F.~{K}upka},
  \bibinfo{author}{E.~{M}undprecht}, \bibinfo{author}{F.~{Z}aussinger},
  \bibinfo{author}{H.~{Grimm-Strele}}, \bibinfo{author}{N.~{H}appenhofer},
\newblock \bibinfo{title}{{S}imulations of stellar convection, pulsation and
  semiconvection},
\newblock in: \bibinfo{editor}{N.~Brummell}, \bibinfo{editor}{A.~Brun},
  \bibinfo{editor}{M.~Miesch}, \bibinfo{editor}{Y.~Ponty} (Eds.),
  \bibinfo{booktitle}{Astrophysical Dynamics: From Stars to Galaxies}, number
  \bibinfo{number}{271} in \bibinfo{series}{Proceedings IAU Symposium}, pp.
  \bibinfo{pages}{179 -- 186}.
\bibitem[{{M}undprecht et~al.(2013){M}undprecht, {M}uthsam, and
  {K}upka}]{MundprechtMuthsamKupka2013}
\bibinfo{author}{E.~{M}undprecht}, \bibinfo{author}{H.~J. {M}uthsam},
  \bibinfo{author}{F.~{K}upka},
\newblock \bibinfo{title}{{M}ultidimensional realistic modelling of
  {C}epheid-like variables. {I}. {E}xtensions of the {ANTARES} code},
\newblock \bibinfo{journal}{MNRAS} \bibinfo{volume}{435} (\bibinfo{year}{2013})
  \bibinfo{pages}{3191 -- 3205}.
\bibitem[{{S}hu and {O}sher(1988)}]{ShuOsher1988}
\bibinfo{author}{C.-W. {S}hu}, \bibinfo{author}{S.~{O}sher},
\newblock \bibinfo{title}{{E}fficient implementation of essentially
  non-oscillatory shock-capturing schemes},
\newblock \bibinfo{journal}{JCP} \bibinfo{volume}{77} (\bibinfo{year}{1988})
  \bibinfo{pages}{439 -- 471}.
\bibitem[{{S}hu(2003)}]{Shu2003}
\bibinfo{author}{C.-W. {S}hu},
\newblock \bibinfo{title}{{H}igh-order {F}inite {D}ifference and {F}inite
  {V}olume {WENO} {S}chemes and {D}iscontinuous {G}alerkin {M}ethods for
  {CFD}},
\newblock \bibinfo{journal}{International Journal of Computational Fluid
  Dynamics} \bibinfo{volume}{17} (\bibinfo{year}{2003}) \bibinfo{pages}{107 --
  118}.
\bibitem[{{M}erriman(2003)}]{Merriman2003}
\bibinfo{author}{B.~{M}erriman},
\newblock \bibinfo{title}{{U}nderstanding the {S}hu--{O}sher {C}onservative
  {F}inite {D}ifference {F}orm},
\newblock \bibinfo{journal}{Journal Of Scientific Computing}
  \bibinfo{volume}{19} (\bibinfo{year}{2003}) \bibinfo{pages}{309 -- 322}.
\bibitem[{{M}uthsam et~al.(2007){M}uthsam, {L\"ow-Baselli}, {O}bertscheider,
  {L}anger, {L}enz, and {K}upka}]{MuthsamLoew-BaselliOberscheideretal2007}
\bibinfo{author}{H.~J. {M}uthsam}, \bibinfo{author}{B.~{L\"ow-Baselli}},
  \bibinfo{author}{C.~{O}bertscheider}, \bibinfo{author}{M.~{L}anger},
  \bibinfo{author}{P.~{L}enz}, \bibinfo{author}{F.~{K}upka},
\newblock \bibinfo{title}{{H}igh--resolution models of solar granulation: the
  two--dimensional case},
\newblock \bibinfo{journal}{MNRAS} \bibinfo{volume}{380} (\bibinfo{year}{2007})
  \bibinfo{pages}{1335 -- 1340}.
\bibitem[{{W}esseling(2001)}]{Wesseling2001}
\bibinfo{author}{P.~{W}esseling}, \bibinfo{title}{{P}rinciples of
  {C}omputational {F}luid {D}ynamics}, volume~\bibinfo{volume}{29} of
  \textit{\bibinfo{series}{Springer Series in Computational Mathematics}},
  \bibinfo{publisher}{Springer}, \bibinfo{address}{Berlin Heidelberg
  {New--York}}, \bibinfo{year}{2001}.
\bibitem[{{F}erziger and {P}eri\'{c}(2002)}]{FerzigerPeric2002}
\bibinfo{author}{J.~H. {F}erziger}, \bibinfo{author}{M.~{P}eri\'{c}},
  \bibinfo{title}{{C}omputational {M}ethods for {F}luid {D}ynamics},
  \bibinfo{publisher}{Springer}, \bibinfo{address}{Berlin},
  \bibinfo{edition}{3rd} edition, \bibinfo{year}{2002}.
\bibitem[{{L}e{V}eque(2004)}]{LeVeque2004}
\bibinfo{author}{R.~J. {L}e{V}eque}, \bibinfo{title}{{F}inite--{V}olume
  {M}ethods for {H}yperbolic {P}roblems}, Cambridge Texts in Applied
  Mathematics, \bibinfo{publisher}{Cambridge University Press},
  \bibinfo{year}{2004}.
\bibitem[{{K}ifonidis and {M}\"uller(2012)}]{KifonidisMueller2012}
\bibinfo{author}{K.~{K}ifonidis}, \bibinfo{author}{E.~{M}\"uller},
\newblock \bibinfo{title}{{O}n multigrid solution of the implicit equations of
  hydrodynamics. {E}xperiments for the compressible {E}uler equations in
  general coordinates},
\newblock \bibinfo{journal}{A\&A} \bibinfo{volume}{544} (\bibinfo{year}{2012})
  \bibinfo{pages}{A47}.
\bibitem[{{T}hompson et~al.(1985){T}hompson, {W}arsi, and {Wayne
  Mastin}}]{ThompsonWarsiMastin1985}
\bibinfo{author}{J.~F. {T}hompson}, \bibinfo{author}{Z.~U.~A. {W}arsi},
  \bibinfo{author}{C.~{Wayne Mastin}}, \bibinfo{title}{{N}umerical {G}rid
  {G}eneration. {F}oundations and {A}pplications},
  \bibinfo{publisher}{North--Holland}, \bibinfo{year}{1985}.
\bibitem[{{E}vans(2002)}]{Evans2002}
\bibinfo{author}{L.~C. {E}vans}, \bibinfo{title}{{P}artial {D}ifferential
  {E}quations}, volume~\bibinfo{volume}{19} of
  \textit{\bibinfo{series}{Graduate Studies in Mathematics}},
  \bibinfo{publisher}{American Mathematical Society}, \bibinfo{edition}{2nd}
  edition, \bibinfo{year}{2002}.
\bibitem[{{C}horin and {M}arsden(1993)}]{ChorinMarsden1993}
\bibinfo{author}{A.~J. {C}horin}, \bibinfo{author}{J.~E. {M}arsden},
  \bibinfo{title}{{A} {M}athematical {I}ntroduction to {F}luid {M}echanics},
  volume~\bibinfo{volume}{4} of \textit{\bibinfo{series}{Texts in Applied
  Mathematics}}, \bibinfo{publisher}{Springer}, \bibinfo{address}{{New--York}
  Berlin Heidelberg}, \bibinfo{edition}{3rd} edition, \bibinfo{year}{1993}.
\bibitem[{{V}isbal and {G}aitonde(2002)}]{VisbalGaitonde2002}
\bibinfo{author}{M.~R. {V}isbal}, \bibinfo{author}{D.~V. {G}aitonde},
\newblock \bibinfo{title}{{O}n the {U}se of {H}igher-{O}rder
  {F}inite-{D}ifference {S}chemes on {C}urvilinear and {D}eforming {M}eshes},
\newblock \bibinfo{journal}{JCP} \bibinfo{volume}{181} (\bibinfo{year}{2002})
  \bibinfo{pages}{155 -- 185}.
\bibitem[{{N}onomura et~al.(2010){N}onomura, {I}izuka, and
  {F}ujii}]{NonomuraIizukaFujii2010}
\bibinfo{author}{T.~{N}onomura}, \bibinfo{author}{N.~{I}izuka},
  \bibinfo{author}{K.~{F}ujii},
\newblock \bibinfo{title}{{F}reestream and vortex preservation properties of
  high-order {WENO} and {WCNS} on curvilinear grids},
\newblock \bibinfo{journal}{Computers \& Fluids} \bibinfo{volume}{39}
  (\bibinfo{year}{2010}) \bibinfo{pages}{197 -- 214}.
\bibitem[{{C}olella et~al.(2011){C}olella, {D}orr, {H}ittinger, and
  {M}artin}]{ColellaDorrHittingerMartin2011}
\bibinfo{author}{P.~{C}olella}, \bibinfo{author}{M.~R. {D}orr},
  \bibinfo{author}{J.~A.~F. {H}ittinger}, \bibinfo{author}{D.~F. {M}artin},
\newblock \bibinfo{title}{{H}igh-order, finite-volume methods in mapped
  coordinates},
\newblock \bibinfo{journal}{JCP} \bibinfo{volume}{230} (\bibinfo{year}{2011})
  \bibinfo{pages}{2952 -- 2976}.
\bibitem[{{D}onat and {M}arquina(1996)}]{DonatMarquina1996}
\bibinfo{author}{R.~{D}onat}, \bibinfo{author}{A.~{M}arquina},
\newblock \bibinfo{title}{{C}apturing {S}hock {R}eflections: {A}n {I}mproved
  {F}lux {F}ormula},
\newblock \bibinfo{journal}{JCP} \bibinfo{volume}{125} (\bibinfo{year}{1996})
  \bibinfo{pages}{42 -- 58}.
\bibitem[{{K}raaijevanger(1991)}]{Kraaijevanger1991}
\bibinfo{author}{J.~F.~B.~M. {K}raaijevanger},
\newblock \bibinfo{title}{{C}ontractivity of {R}unge-{K}utta methods},
\newblock \bibinfo{journal}{BIT} \bibinfo{volume}{31} (\bibinfo{year}{1991})
  \bibinfo{pages}{482 -- 528}.
\bibitem[{{K}upka et~al.(2012){K}upka, {H}appenhofer, {H}igueras, and
  {K}och}]{KupkaHappenhoferHiguerasKoch2012}
\bibinfo{author}{F.~{K}upka}, \bibinfo{author}{N.~{H}appenhofer},
  \bibinfo{author}{I.~{H}igueras}, \bibinfo{author}{O.~{K}och},
\newblock \bibinfo{title}{{T}otal-variation-diminishing implicit--explicit
  {R}unge--{K}utta methods for the simulation of double-diffusive convection in
  astrophysics},
\newblock \bibinfo{journal}{JCP} \bibinfo{volume}{231} (\bibinfo{year}{2012})
  \bibinfo{pages}{3561 -- 3586}.
\bibitem[{{M}iczek(2013)}]{Miczek2013}
\bibinfo{author}{F.~{M}iczek}, \bibinfo{title}{{S}imulation of low {M}ach
  number astrophysical flows}, Ph.D. thesis, TU M\"unchen,
  \bibinfo{year}{2013}.
\bibitem[{{L}iska and {W}endroff(2003)}]{LiskaWendroff2003}
\bibinfo{author}{R.~{L}iska}, \bibinfo{author}{B.~{W}endroff},
\newblock \bibinfo{title}{{C}omparions of {S}everal {D}ifference {S}chemes on
  1{D} and 2{D} {T}est {P}roblems for the {E}uler {E}quations},
\newblock \bibinfo{journal}{SIAM Journal on Scientific Computing}
  \bibinfo{volume}{25} (\bibinfo{year}{2003}) \bibinfo{pages}{1 -- 30}.
  \bibinfo{note}{Available at
  \url{http://www-troja.fjfi.cvut.cz/~liska/CompareEuler/compare8/}}.
\bibitem[{{S}od(1978)}]{Sod1978}
\bibinfo{author}{G.~A. {S}od},
\newblock \bibinfo{title}{{A} survey of several finite difference methods for
  systems of nonlinear hyperbolic conservation laws},
\newblock \bibinfo{journal}{JCP} \bibinfo{volume}{27} (\bibinfo{year}{1978})
  \bibinfo{pages}{1 -- 31}.
\bibitem[{{Y}ee et~al.(1999){Y}ee, {S}andham, and
  {D}jomehri}]{YeeSandhamDjomehri1999}
\bibinfo{author}{H.~{Y}ee}, \bibinfo{author}{N.~{S}andham},
  \bibinfo{author}{M.~{D}jomehri},
\newblock \bibinfo{title}{{L}ow-{D}issipative {H}igh-{O}rder
  {S}hock-{C}apturing {M}ethods {U}sing {C}haracteristic-{B}ased {F}ilters},
\newblock \bibinfo{journal}{JCP} \bibinfo{volume}{150} (\bibinfo{year}{1999})
  \bibinfo{pages}{199 -- 238}.
\bibitem[{{Z}hang et~al.(2011){Z}hang, {Z}hang, and {S}hu}]{ZhangZhangShu2011}
\bibinfo{author}{R.~{Z}hang}, \bibinfo{author}{M.~{Z}hang},
  \bibinfo{author}{C.-W. {S}hu},
\newblock \bibinfo{title}{{O}n the {O}rder of {A}ccuracy and {N}umerical
  {P}erformance of {T}wo {C}lasses of {F}inite {V}olume {WENO} {S}chemes},
\newblock \bibinfo{journal}{Commun. Comput. Phys.} \bibinfo{volume}{9}
  (\bibinfo{year}{2011}) \bibinfo{pages}{807 -- 827}.
\bibitem[{{C}asper and {A}tkins(1993)}]{CasperAtkins1993}
\bibinfo{author}{J.~{C}asper}, \bibinfo{author}{H.~L. {A}tkins},
\newblock \bibinfo{title}{{A} {F}inite-{V}olume {H}igh-{O}rder {ENO} {S}cheme
  for {T}wo-{D}imensional {H}yperbolic {S}ystems},
\newblock \bibinfo{journal}{JCP} \bibinfo{volume}{106} (\bibinfo{year}{1993})
  \bibinfo{pages}{62 -- 76}.
\bibitem[{{K}etcheson et~al.(2009){K}etcheson, {M}acdonald, and
  {G}ottlieb}]{KetchesonMacdonaldGottlieb2009}
\bibinfo{author}{D.~I. {K}etcheson}, \bibinfo{author}{C.~B. {M}acdonald},
  \bibinfo{author}{S.~{G}ottlieb},
\newblock \bibinfo{title}{{O}ptimal implicit strong stability preserving
  {Runge--Kutta} methods},
\newblock \bibinfo{journal}{Applied Numerical Mathematics} \bibinfo{volume}{59}
  (\bibinfo{year}{2009}) \bibinfo{pages}{373 -- 392}.
\bibitem[{{K}watra et~al.(2009){K}watra, {S}u, {G}retarsson, and
  {F}edkiw}]{KwatraSuGretarssonFedkiw2009}
\bibinfo{author}{N.~{K}watra}, \bibinfo{author}{J.~{S}u},
  \bibinfo{author}{J.~T. {G}retarsson}, \bibinfo{author}{R.~{F}edkiw},
\newblock \bibinfo{title}{{A} method for avoiding the acoustic time step
  restriction in compressible flow},
\newblock \bibinfo{journal}{JCP} \bibinfo{volume}{228} (\bibinfo{year}{2009})
  \bibinfo{pages}{4146 -- 4161}.
\bibitem[{{V}inokur(1974)}]{Vinokur1974}
\bibinfo{author}{M.~{V}inokur},
\newblock \bibinfo{title}{{C}onservation {E}quations of {G}asdynamics in
  {C}urvilinear {C}oordinate {S}ystems},
\newblock \bibinfo{journal}{JCP} \bibinfo{volume}{14} (\bibinfo{year}{1974})
  \bibinfo{pages}{105 -- 125}.
\bibitem[{{T}annehill et~al.(1997){T}annehill, {A}nderson, and
  {P}letcher}]{TannehillAndersonPletcher1997}
\bibinfo{author}{J.~C. {T}annehill}, \bibinfo{author}{D.~A. {A}nderson},
  \bibinfo{author}{R.~H. {P}letcher}, \bibinfo{title}{{C}omputational {F}luid
  {M}echanics and {H}eat {T}ransfer}, \bibinfo{publisher}{Taylor \& Francis},
  \bibinfo{year}{1997}.

\end{thebibliography}
\end{document}